\documentclass[lettersize,journal]{IEEEtran}
\usepackage{amssymb}
\usepackage{amsmath}
\usepackage{amsthm}
\usepackage{amsfonts}
\usepackage{algorithm}
\usepackage{algorithmic}
\usepackage{array}
\usepackage{bbding}
\usepackage{bigints}
\usepackage{booktabs}
\usepackage{cite}
\usepackage{cleveref}
\usepackage{color}
\usepackage{diagbox}
\usepackage{epsfig,latexsym}
\usepackage{epstopdf}
\usepackage{graphicx}
\usepackage{fancyhdr}
\usepackage{float}
\usepackage{flushend}
\usepackage{indentfirst}
\usepackage{lastpage}
\usepackage{makecell}
\usepackage{mathtools}
\usepackage{multirow}
\usepackage{pifont}
\usepackage{psfrag}
\usepackage{setspace}
\usepackage{stfloats}
\usepackage{subfloat}
\usepackage{times}
\usepackage{arydshln}
\usepackage{textcomp}
\usepackage{url}
\usepackage{verbatim}
\usepackage[caption=false,font=normalsize,labelfont=sf,textfont=sf]{subfig}

\newcommand{\cmark}{\ding{52}}

\theoremstyle{remark}
\newtheorem{theorem}{\quad \textbf{Theorem}}
\newtheorem{lemma}{\quad \textbf{Lemma}}

\newtheorem{corollary}{\quad \textbf{Corollary}}

\allowdisplaybreaks[4]

\begin{document}
\title{Performance Analysis of Reconfigurable Holographic Surfaces in the Near-Field Scenario of Cell-Free Networks Under Hardware Impairments}

\author{Qingchao Li, \textit{Graduate Student Member, IEEE}, Mohammed El-Hajjar, \textit{Senior Member, IEEE},\\ Yanshi Sun, \textit{Member, IEEE}, and Lajos Hanzo, \textit{Life Fellow, IEEE}

\thanks{Lajos Hanzo would like to acknowledge the financial support of the Engineering and Physical Sciences Research Council projects EP/W016605/1, EP/X01228X/1, EP/Y026721/1 and EP/W032635/1 as well as of the European Research Council's Advanced Fellow Grant QuantCom (Grant No. 789028). \textit{(Corresponding author: Lajos Hanzo.)}

Qingchao Li, Mohammed El-Hajjar and Lajos Hanzo are with the School of Electronics and Computer Science, University of Southampton, Southampton SO17 1BJ, U.K. (e-mail: Qingchao.Li@soton.ac.uk; meh@ecs.soton.ac.uk; lh@ecs.soton.ac.uk).

Yanshi Sun is with the School of Computer Science and Information Engineering, Hefei University of Technology, Hefei, 230009, China. (email:sys@hfut.edu.cn).}}

\maketitle

\begin{abstract}
We propose a hybrid beamforming architecture for near-field reconfigurable holographic surfaces (RHS) harnessed in cell-free networks. Specifically, the holographic beamformer of each base station (BS) is designed for maximizing the channel gain based on the local channel state information (CSI). By contrast, the digital beamformer at the central processing unit is designed based on the minimum mean squared error criterion. Furthermore, the near-field spectral efficiency of the RHS in cell-free networks is derived theoretically by harnessing the popular stochastic geometry approach. We consider both the phase shift error (PSE) at the RHS elements and the hardware impairment (HWI) at the radio frequency (RF) chains of the transceivers. Furthermore, we theoretically derive the asymptotic capacity bound, when considering an infinite physical size for the RHS in the near-field channel model. The theoretical analysis and simulation results show that the PSE at the RHS elements and the HWI at the RF chains of transceivers limit the spectral efficiency in the high signal-to-noise ratio region. Moreover, we show that the PSE at the RHS elements and the HWI at the RF chains of BSs can be compensated by increasing the number of BSs. Finally, we also demonstrate that the ergodic spectral efficiency based on the near-field channel model is higher than that based on the far-field channel model assumption.
\end{abstract}
\begin{IEEEkeywords}
Reconfigurable holographic surfaces (RHS), near-field, cell-free network, stochastic geometry, phase shift error (PSE), hardware impairment (HWI).
\end{IEEEkeywords}

\section{Introduction}
\IEEEPARstart{D}{riven} by the every-increasing demand for data transmission in mobile networks, sophisticated technologies, such as millimeter wave (mmWave) communications, massive multiple-input and multiple-output (MIMO) systems, and ultra-dense networks (UDN) have been rolled out across the globe~\cite{gao2015mmwave}. Hence, at the time of writing, research is also well under way for the exploration of next-generation communications~\cite{huang2022general}.

The emerging reconfigurable holographic surface (RHS) is a promising solution for the future wireless networks~\cite{deng2023reconfigurable}. In contrast to intelligent reflecting surfaces (IRS), which play the role of a relay conceived for reconfiguring the propagation environment~\cite{tang2022path}, the RHS acts as a reconfigurable antenna array at base stations (BSs). It has the benefit of a spatially continuous electromagnetic aperture capable of creating pencil beams to approach the ultimate capacity limit of wireless channels~\cite{gong2023holographic}. Specifically, the RHS is a spatially finite metasurface constructed by numerous reconfigurable radiation elements, essentially realizing a software-controlled antenna. Compared to the conventional massive MIMO paradigm that employs a large number of high-complexity and energy-hungry radio frequency (RF) chains, the RHS has a reduced number of RF chains. Its beamforming pattern is formed by appropriately adjusting the phase and amplitude of the reconfigurable elements based on the holographic principle.

The RHS has been widely investigated, including its channel modeling~\cite{pizzo2020spatially}, channel estimation~\cite{demir2022channel}, \cite{an2023tutorial_part_I}, near-filed communications~\cite{an2023toward}, \cite{an2023tutorial_part_II} and its applications in sensing~\cite{zhang2022holographic}. As for its beamforming techniques, Wei \textit{et al.}~\cite{wei2022multi} proposed a low-complexity zero-forcing (ZF) based beamforming architecture relying on Neumann series expansion to replace the matrix inversion operation in the downlink of multi-user RHS-based MIMO communications. Moreover, both user-cluster based precoding method and a two-layer precoding method were proposed in~\cite{wei2023tri} for mitigating the cross-polarization and inter-user interferences in tri-polarized holographic MIMO schemes. In~\cite{deng2022hdma}, a novel type of space-division multiple access (SDMA), termed as holographic-pattern division multiple access (HDMA), was proposed based on the RHS architecture. The theoretical analysis and simulation results showed that the HDMA scheme outperforms the traditional SDMA arrangement in terms of both its cost-efficiency and sum-rate. Moreover, in~\cite{zhang2023pattern}, Zhang \textit{et al.} employed the HDMA technique and designed continuous-aperture based MIMO patterns, where the continuous pattern functions were transformed into their projection on finite orthogonal bases. Then the block coordinate descent (BCD) method was employed for solving the associated sum-rate maximization problem. Deng \textit{et al.}~\cite{deng2022reconfigurable_twc}, \cite{deng2021reconfigurable_tvt} proposed an amplitude-controlled RHS architecture, where a hybrid beamforming scheme was employed for maximizing the achievable sum-rate. Specifically, the associated digital beamformer was designed based on the ZF method, while the holographic beamformer was realized by appropriately adjusting the amplitude of the reference wave radiated in order to form holographic patterns in the desired beam directions. It was shown in~\cite{deng2022reconfigurable_twc}, \cite{deng2021reconfigurable_tvt} that the RHS-assisted hybrid beamformer promises higher sum-rate than the conventional phased array based MIMO systems for the same physical size, while imposing a reduced hardware cost. Moreover, the holographic beamformer based on the discrete amplitude-controlled RHS elements and the corresponding amplitude discretization effect were investigated in~\cite{hu2022holographic}. An \textit{et al.} in~\cite{an2023stacked} proposed a stacked intelligent metasurface (SIM) based architecture, where multiple holographic surfaces are stacked at the transceiver. To maximize the achievable sum-rate, the phase shifts associated with all the metasurface layers of the SIM were optimized based on the gradient descent algorithm. By contrast, the digital beamformer at the RF chains was designed based on the singular value decomposition (SVD) method. The theoretical analysis and simulation results showed that the achievable sum-rate performance can be improved upon increasing the number of RHS layers. For mitigating the channel state information (CSI) acquisition overhead, Wu \textit{et al.}~\cite{wu2023two} proposed a two-time scale beamforming scheme. Specifically, the holographic beamformer was designed based on the slow-changed statistical CSI. Then the instantaneous CSI of the equivalent channel links was estimated and exploited for designing the digital precoding matrix. Furthermore, the asymptotic capacity of the RHS in the near-field channel model is investigated in~\cite{zeng2022reconfigurable}. It was shown that the RHS consumes less power than the phased array, when the user equipment (UE) is in the far field of the BS antenna.

However, the above RHS solutions have the following limitations.

Firstly, the beamforming techniques in the above RHS solutions were tailored for traditional cellular networks, where the cell-edge UEs suffer from low-reliability reception due to the high path-loss and inter-cell interference. The cell-free network concept has emerged as a promising technique for achieving an increased higher rate as well as near-uniform load-balancing, where a central BS is replaced by multiple distributed BSs to serve multiple UEs~\cite{chen2023improving}. As a benefit of the reduced average distance between the UE and its nearest BS as well as the coordination among BSs, the transmission reliability can be significantly improved~\cite{bjornson2020scalable}. In~\cite{zhang2021local}, Zhang \textit{et al.} analyzed the uplink spectral efficiency of various scalable combining methods, including full-pilot zero-forcing (FZF), partial FZF (PFZF), protective weak PFZF (PWPFZF), and local regularized ZF (LRZF) schemes, in massive MIMO systems by exploiting the  associated channel statistics. Numerical results show that LRZF provides the highest spectral efficiency. Ma \textit{et al.}~\cite{ma2023cooperative} proposed a novel partially-connected cell-free massive MIMO (P-CF-mMIMO) framework for reconfigurable intelligent surface (RIS) assisted systems, where a limited number of communication links exist among the users and BSs for the sake of reducing the communication costs. They showed that the P-CF-mMIMO systems achieve an improved performance or communication costs compromise, by selecting an appropriate network connection ratio. To reduce the overhead required for CSI-sharing between BSs, a distributed algorithm can be employed based on the maximum ratio transmission (MRT) or the maximum ratio combining (MRC) criteria, albeit at the cost of a performance degradation. Specifically, in the distributed optimization algorithm of the cell-free system, the cooperation between BSs is promising in terms of harnessing parallel computing resources and achieving almost the same data rate as the centralized algorithm~\cite{xu2024algorithm}. Furthermore, as shown in~\cite{an2023stacked_arxiv}, the cell-free scenario can be beneficially conceived with the SIM-aided MIMO systems to reduce both the energy consumption and the processing delay.

Secondly, the beamforming design and performance analysis in~\cite{wei2022multi}, \cite{wei2023tri}, \cite{deng2022hdma}, \cite{zhang2023pattern}, \cite{deng2022reconfigurable_twc}, \cite{deng2021reconfigurable_tvt}, \cite{hu2022holographic}, \cite{an2023stacked}, \cite{wu2023two}, \cite{zeng2022reconfigurable} are based on the assumption of ideal phase shift configuration at the RHS elements. However, the phase shift error (PSE) is inevitable at the reconfigurable surfaces in practical hardware, resulting in significant performance degradation~\cite{badiu2019communication}. Moreover, the hardware impairments (HWI) at the RF chains of transceivers are also inevitable, which inflict performance degradations as well~\cite{papazafeiropoulos2021intelligent}.

To deal with the above issues, in this paper we propose a hybrid beamformer for the RHS-based cell-free networks and derive the corresponding theoretical spectral efficiency upper bound in the near-field, in the face of the realistic PSE at the RHS elements and practical HWIs at the RF chains of the transceivers. Our contributions in this paper are as follows:
\begin{itemize}
  \item To reduce the channel estimation overhead, we propose a hybrid beamforming architecture for RHS-based cell-free networks in the face of imperfect hardware quality of the phase shifters at the RHS elements and of the RF chains at the transceivers. Specifically, the holographic beamformer of the distributed BSs is designed based on the local CSI by adjusting the phase shifts of the RHS for maximizing the power gain, while the digital beamformer at the central processing unit (CPU) is designed by relying on the minimum mean squared error (MMSE) method based on the overall CSI shared by the distributed BSs for mitigating the inter-user interference. We also take into account the PSE of the RHS elements and the HWIs of the RF chains.
  \item The theoretical ergodic spectral efficiency of RHS-based cell-free networks is derived, in the face of the PSE and HWIs, by leveraging the popular stochastic geometry approach. The theoretical analysis shows that the spectral efficiency is limited by the PSE of the RHS elements in the high signal-to-noise ratio (SNR) region, but it may be compensated by increasing the number of BSs. Similarly, the spectral efficiency is limited by the HWIs of the transceivers in the high-SNR region. Moreover, increasing the number of BSs is capable of compensating the HWIs of the RF chains at the BSs, but it cannot compensate for the HWIs of the RF chains at the UEs.
  \item  Furthermore, the asymptotic approximate capacity bound is derived for the infinite physical dimension of the RHS in a near-field scenario. We show that the ergodic spectral efficiency based on the near-field channel model is higher than that based on the far-field channel model assumption.
\end{itemize}

Finally, Table~\ref{Table_literature} explicitly contrasts our contributions to the literature~\cite{wei2022multi}, \cite{wei2023tri}, \cite{deng2022hdma}, \cite{zhang2023pattern}, \cite{deng2022reconfigurable_twc}, \cite{deng2021reconfigurable_tvt}, \cite{hu2022holographic}, \cite{an2023stacked}, \cite{wu2023two}, \cite{zeng2022reconfigurable}.

\begin{table*}
\footnotesize
\begin{center}
\caption{Novelty comparison of our paper to the existing RHS techniques in literature~\cite{wei2022multi}, \cite{wei2023tri}, \cite{deng2022hdma}, \cite{zhang2023pattern}, \cite{deng2022reconfigurable_twc}, \cite{deng2021reconfigurable_tvt}, \cite{hu2022holographic}, \cite{an2023stacked}, \cite{wu2023two}, \cite{zeng2022reconfigurable}.}
\label{Table_literature}
\begin{tabular}{*{12}{l}}
\hline
     & \makecell[c]{Our paper} & \cite{wei2022multi} & \cite{wei2023tri} & \cite{deng2022hdma} & \cite{zhang2023pattern} & \cite{deng2022reconfigurable_twc} & \cite{deng2021reconfigurable_tvt} & \cite{hu2022holographic} & \cite{an2023stacked} & \cite{wu2023two} & \cite{zeng2022reconfigurable}\\
\hline
    Holographic beamforming & \makecell[c]{\cmark} & \makecell[c]{\cmark} & \makecell[c]{\cmark} & \makecell[c]{\cmark} & \makecell[c]{\cmark} & \makecell[c]{\cmark} & \makecell[c]{\cmark} & \makecell[c]{\cmark} & \makecell[c]{\cmark} & \makecell[c]{\cmark} & \makecell[c]{\cmark} \\
\hdashline
    Performance analysis & \makecell[c]{\cmark} & \makecell[c]{\cmark} & \makecell[c]{\cmark} & \makecell[c]{\cmark} & \makecell[c]{\cmark} &  &  & \makecell[c]{\cmark} & \makecell[c]{\cmark} &  & \makecell[c]{\cmark} \\
\hdashline
    Near-field channel model & \makecell[c]{\cmark} &  & \makecell[c]{\cmark} & \makecell[c]{\cmark} & \makecell[c]{\cmark} &  &  &  & \makecell[c]{\cmark} & & \makecell[c]{\cmark} \\
\hdashline
    Cell-free networks & \makecell[c]{\cmark} &  &  &  &  &  &  &  &  &  & \\
\hdashline
    RHS phase shift error & \makecell[c]{\cmark} &  &  &  &  &  &  &  &  &  & \\
\hdashline
    Transceivers hardware impairments & \makecell[c]{\cmark} &  &  &  &  &  &  &  &  &  & \\
\hline
\end{tabular}
\end{center}
\end{table*}

The rest of this paper is organized as follows. In Section~\ref{System_Model}, we present the system model, while Section~\ref{Beamforming_Design} highlights our hybrid beamformer design. Our theoretical analysis is presented in Section~\ref{Theoretical_Analysis}, complemented by our simulation results in Section~\ref{Numerical_and_Simulation_Results}. Finally, we conclude in Section~\ref{Conclusion}.

\textit{Notations:} Vectors and matrices are denoted by boldface lower and upper case letters, respectively; $(\cdot)^{\mathrm{T}}$, $(\cdot)^{\dag}$ and $(\cdot)^{\mathrm{H}}$ represent the operation of transpose, conjugate and hermitian transpose, respectively; $\odot$ represents the Hadamard product operation; $|a|$ represents the amplitude of the complex scalar $a$; $\|\mathbf{a}\|$ represents the Euclidean norm of the vector $\mathbf{a}$; $\mathbb{C}^{m\times n}$ denotes the space of $m\times n$ complex-valued matrices; $\mathbf{I}_{N}$ represents the $N\times N$ identity matrix; $\mathbf{Diag}\{a_1,a_2,\cdots,a_N\}$ denotes a diagonal matrix with the diagonal elements being the elements of $a_1,a_2,\cdots,a_N$ in order; $a_n$ represents the $n$th element in the vector $\mathbf{a}$; $\mathcal{CN}(\boldsymbol{\mu},\mathbf{\Sigma})$ is a circularly symmetric complex Gaussian random vector with the mean $\boldsymbol{\mu}$ and the covariance matrix $\mathbf{\Sigma}$; $|\Phi|$ represents the cardinal number of the set $\Phi$; $\mathbf{C}_{\mathbf{a}\mathbf{b}}$ represents the correlation matrix between the vectors $\mathbf{a}$ and $\mathbf{b}$, i.e. $\mathbf{C}_{\mathbf{a}
\mathbf{b}}=\mathbb{E}[\mathbf{a}\mathbf{b}^{\mathrm{H}}]$.

\section{System Model}\label{System_Model}
In this section, we first describe our proposed reconfigurable holographic surface based cell-free network, followed by its near-field channel model.

\begin{figure*}[!t]
    \centering
    \includegraphics[width=5in]{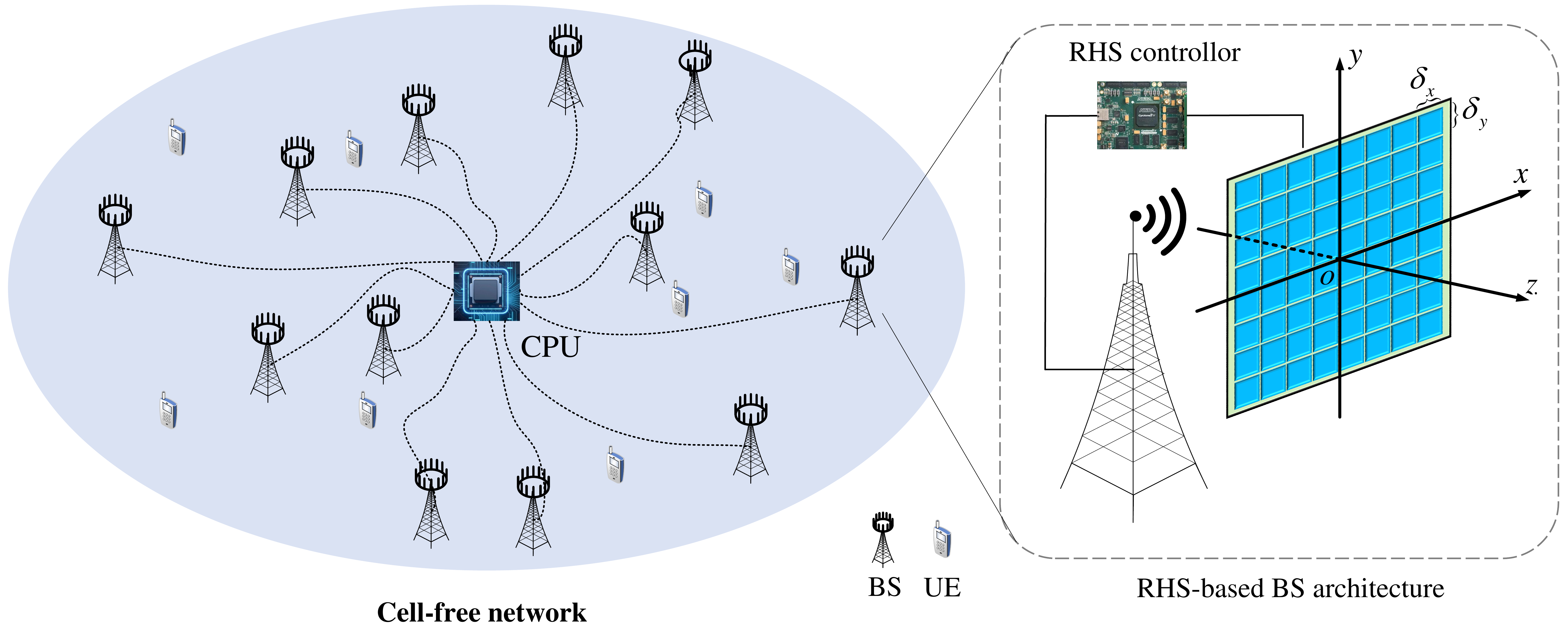}
    \caption{System model of reconfigurable holographic surfaces-based cell-free network.}\label{Fig_system_model_cell_free}
\end{figure*}

The system model of the RHS-based cell-free network is shown in Fig.~\ref{Fig_system_model_cell_free}. In contrast to the conventional cellular network where a central BS supports the UEs in the cell, the cell-free networks relies on multiple distributed BSs which cooperatively serve the UEs. We focus on the uplink operation, where a set of BSs supports $K$ single-antenna UEs. We denote the set of BS locations as $\Phi$. Referring to~\cite{haenggi2012stochastic}, \cite{haenggi2009stochastic}, the distribution of BSs across the two-dimensional Euclidean plane can be modelled by a homogeneous Poisson point process (PPP) with density $\eta$, measured in terms of the number of BSs per $\mathrm{m}^2$. Furthermore, the RHSs are at the height of $H$ and have a random orientation following a uniform distribution. To reduce the CSI estimation overhead, we employ hybrid beamforming, where the holographic beamformer of each distributed BS is designed based on the local CSI. Then the CPU fuses the signals gleaned from all distributed BSs via the fronthaul and designs the digital beamformer based on the overall CSI received from all BSs to detect the UEs' desired information. Furthermore, due to the hardware imperfections of practical systems, having HWIs at the RF chains~\cite{bjornson2017massive}, \cite{bjornson2014massive} and PSE at the RHS elements~\cite{badiu2019communication}, \cite{qian2020beamforming} are inevitable. In this paper, we consider both of these non-ideal characteristics and analyse their effects on the system performance.

\subsection{RHS-based BS Architecture}
In contrast to the IRS deployed in a channel environment to play the role of a relay conceived for reconfiguring the propagation environment, the RHS is deployed at the BSs to act as reconfigurable antenna arrays. To avoid having a large number energy-hungry RF chains to support energy-efficient communications, where a RHS illuminated by a single RF chain can be utilized at the cell-free BSs~\cite{zeng2022reconfigurable}, \cite{zhang2022intelligent}\footnote{To unveil the theoretically achievable rate limit of the RHS in the near-field scenario of cell-free networks, we assume that each user is equipped with a single antenna. Practically, each user may be equipped with multiple antennas to acquire higher diversity gain. Note that our methodology is also applicable to the case of multiple antennas at the users, when the beamformers at the RHS and the users are alteratively optimized.}. As shown in Fig.~\ref{Fig_system_model_cell_free}, the $xoy$ plane coincides with the RHS and the origin $o$ is located at the center of the RHS. We assume that a total of $N=N_x\times N_y$ RHS elements are compactly placed in a uniform rectangular planar array, with $N_x$ elements in the $x$ axis direction and $N_y$ elements in the $y$ axis direction. Each RHS element has the size of $A=\delta_x\times\delta_y$.

In BS-$l$ ($l=1,2,\cdots,L$), we denote the coefficient of the $n$th element by $\Gamma_n^{(l)}\mathrm{e}^{\jmath\theta_n^{(l)}}$, where $L=|\Phi|$ represent the number of BSs while $\Gamma_n^{(l)}\in[0,1]$ and $\theta_n^{(l)}\in(-\pi,\pi]$ represent the appropriately configured amplitude and phase shift of the $n$th element, respectively. We set $\Gamma_n^{(l)}=1$ to realize full energy transfer, and hence the response of the RHS is given by
\begin{align}\label{System_Model_1}
    \notag\mathbf{\Theta}^{(l)}=&\mathbf{Diag}\left\{\mathrm{e}^{\jmath\theta_1^{(l)}},
    \mathrm{e}^{\jmath\theta_2^{(l)}},\cdots,\mathrm{e}^{\jmath\theta_N^{(l)}}\right\}\\
    =&\mathbf{Diag}\left\{\mathrm{e}^{\jmath\left(\overline{\theta}_1^{(l)}
    +\widetilde{\theta}_1^{(l)}\right)},
    \mathrm{e}^{\jmath\left(\overline{\theta}_2^{(l)}+\widetilde{\theta}_2^{(l)}\right)},\cdots,
    \mathrm{e}^{\jmath\left(\overline{\theta}_N^{(l)}+\widetilde{\theta}_N^{(l)}\right)}\right\},
\end{align}
where $\theta_n^{(l)}=\overline{\theta}_n^{(l)}+\widetilde{\theta}_n^{(l)}$ with $\overline{\theta}_n^{(l)}$ being the expected phase shift value of the $n$th RHS element, while $\widetilde{\theta}_n^{(l)}$ represents the phase error due to the realistic RHS hardware imperfection. The phase error $\widetilde{\theta}_n^{(l)}$ obeys identically and independently distributed (i.i.d.) random variables having the mean of 0, and it may also be modelled by the von-Mises distribution or the uniform distribution~\cite{badiu2019communication,
qian2020beamforming}. These may be represented as $\widetilde{\theta}_n^{(l)}\sim\mathcal{VM}(0,\varpi_\text{p})$ and $\widetilde{\theta}_n^{(l)}
\sim\mathcal{UF}(-\iota_\text{p},\iota_\text{p})$, respectively, where $\varpi_\text{p}$ is the concentration parameter of the von-Mises distributed variables and $(-\iota_\text{p},
\iota_\text{p})$ is the support interval of the uniformly distributed variables. Although the exact values of $\widetilde{\theta}_n^{(l)}$ cannot be obtained, we can exploit the statistical information for beamforming designs. For BS-$l$, we represent the desired phase shift matrix by $\overline{\mathbf{\Theta}}^{(l)}=\mathbf{Diag}
\{\mathrm{e}^{\jmath\overline{\theta}_1^{(l)}},\mathrm{e}^{\jmath\overline{\theta}_2^{(l)}},
\cdots,\mathrm{e}^{\jmath\overline{\theta}_N^{(l)}}\}$.

\subsection{Near-field Channel Model}
As shown in Fig.~\ref{Fig_system_model_cell_free}, the equivalent channel spanning from the UEs to the RF chains at the BSs is composed of the channel links impinging from the UEs to the RHS, the RHS beamforming matrices and the channel links spanning from the RHS to the RF chains at the BSs.

\subsubsection{Channel links from the RHS to the RF chains at the BSs}
For each BS, we assume that the RF chain is located on the normal of the RHS through the origin with the distance of $d_0$. Hence the coordinate of the RF chain is $\mathbf{r}=(0,0,-d_0)^\mathrm{T}$. We denote the position of the geometrical center of the $n$th RHS element as $\mathbf{p}_n=(x_n,y_n,0)^\mathrm{T}$. In contrast to the popular IRS which is deployed in the channel environment, the RHS is part of the BS and it is illumined by an RF module in its vicinity. Thus, referring to~\cite{an2023stacked}, \cite{zeng2022reconfigurable}, \cite{an2023stacked_arxiv}, only the signal radiated from the RHS can be received by the RF chain. Furthermore, we represent the channel link spanning from the RHS to the RF chain as $\mathbf{f}^{\mathrm{H}}\in\mathbb{C}^{1\times N}$, with the response corresponding to the $n$th RHS element given by \cite{zeng2022reconfigurable}
\begin{align}\label{System_Model_2}
    f_n^{\dag}=\sqrt{\varsigma_n}
    \mathrm{e}^{-\jmath\frac{2\pi}{\lambda}\left\|\mathbf{r}-\mathbf{p}_n\right\|},
\end{align}
where $\lambda$ is the carrier wavelength, $x_n$ and $y_n$ are the coordinates of the geometrical center of the $n$th RHS element on the $x$-axis and $y$-axis respectively, and $\varsigma_n$ is the power gain of the link between the $n$th RHS element and the RF chain represented as
\begin{align}\label{System_Model_3}
    \notag\varsigma_n=&\int_{x_n-\frac{\delta_x}{2}}^{x_n+\frac{\delta_x}{2}}
    \int_{y_n-\frac{\delta_y}{2}}^{y_n+\frac{\delta_y}{2}}
    \frac{2\left(\alpha+1\right)\left(\frac{d_0}{\sqrt{d_0^2+x^2+y^2}}\right)^{\alpha+1}}
    {4\pi\left(d_0^2+x^2+y^2\right)}\mathrm{d}x\mathrm{d}y\\
    =&\int_{x_n-\frac{\delta_x}{2}}^{x_n+\frac{\delta_x}{2}}
    \int_{y_n-\frac{\delta_y}{2}}^{y_n+\frac{\delta_y}{2}}
    \frac{2\left(\alpha+1\right)d_0^{\alpha+1}}
    {4\pi\left(d_0^2+x^2+y^2\right)^{\frac{\alpha+3}{2}}}\mathrm{d}x\mathrm{d}y,
\end{align}
where $2(\alpha+1)$ is the gain of the RF chain.

\subsubsection{Channel links from the UEs to the RHS}
When the communication distance between transceivers is shorter than the Rayleigh distance, which is formulated by $\frac{2D^2}{\lambda}$ with $D$ being the physical size of the receiver, the electromagnetic waves impinging on the receiver must be accurately modeled as spherical waves~\cite{cui2022near}. Given the short range high-frequency communications in the cell-free network considered, as well as the large size of the reconfigurable holographic surface, we employ the near-field model for accurately characterizing the channel response.

In our theoretical study on the fundamental performance limits, we can assume that the signals from the UEs to the BSs go through a basic free-space line-of-sight (LoS) propagation, which was widely employed for the near-field channel model, as seen in~\cite{deng2022hdma}, \cite{zeng2022reconfigurable}, \cite{lu2021communicating}, \cite{lu2021near}. We denote the channel vector spanning from the UE-$k$ to the RHS at BS-$l$ as $\mathbf{g}^{(l,k)}\in\mathbb{C}^{N\times1}$, with the response corresponding to the $n$th RHS element given by~\cite{lu2021communicating}, \cite{lu2021near}
\begin{align}\label{System_Model_4}
    g_n^{(l,k)}=\sqrt{\beta_n^{(l,k)}}\mathrm{e}^{-\jmath\frac{2\pi}{\lambda}
    \left\|\mathbf{q}^{(l,k)}-\mathbf{p}_n\right\|},
\end{align}
where $\mathbf{q}^{(l,k)}$ is the position of UE-$k$ in the Cartesian coordinate of BS-$l$, and $\beta_n^{(l,k)}$ is the channel's power gain between UE-$k$ and the $n$th RHS element at BS-$l$, given by
\begin{align}\label{System_Model_5}
    \notag\beta_n^{(l,k)}=&\iiint_{\mathcal{D}_n}
    \frac{\sin\psi_\mathbf{t}^{(l,k)}}{4\pi\left\|\mathbf{q}^{(l,k)}
    -\mathbf{t}\right\|^2}\mathrm{d}\mathbf{t}\\
    \overset{(a)}=&\iiint_{\mathcal{D}_n}\frac{\left\|\mathbf{q}^{(l,k)}\right\|\cdot
    \sin\psi^{(l,k)}}{4\pi\left\|\mathbf{q}^{(l,k)}-\mathbf{t}\right\|^3}\mathrm{d}\mathbf{t}.
\end{align}
In (\ref{System_Model_5}) the integration interval of the coordinate interval of the $n$th RHS element is
\begin{align}
    \notag\mathcal{D}_n=&\left\{\left(x,y,0\right)^\mathrm{T}:x_n-\frac{\delta_x}{2}<x\leq x_n+\frac{\delta_x}{2},\right.\\
    &\left.y_n-\frac{\delta_y}{2}<y\leq y_n+\frac{\delta_y}{2}\right\},
\end{align}
$\psi_\mathbf{t}^{(l,k)}$ denotes the angle between the vector $\mathbf{q}^{(l,k)}-\mathbf{t}$ and the $xoy$ plane of Fig.~\ref{Fig_system_model_cell_free} in the Cartesian coordinate of BS-$l$, while $\psi^{(l,k)}$ denotes the angle between the vector $\mathbf{q}^{(l,k)}$ and the $xoy$ plane in the Cartesian coordinate of BS-$l$. Still referring to (\ref{System_Model_5}), (a) is based on $\|\mathbf{q}^{(l,k)}-\mathbf{t}\|\cdot\sin\psi_\mathbf{t}^{(l,k)}
=\|\mathbf{q}^{(l,k)}\|\cdot\sin\psi^{(l,k)}$ for all $\mathbf{t}\in\mathcal{D}_n$. In practical systems, since the size of each RHS element is on the wavelength scale, the channel's power gain variation from UE-$k$ to different points belonging to $\mathcal{D}_n$ is negligible. Therefore, the channel's power gain between UE-$k$ and the $n$th RHS element at BS-$l$ can be approximated as
\begin{align}\label{System_Model_6}
    \beta_n^{(l,k)}\approx\frac{A\left\|\mathbf{q}^{(l,k)}\right\|\sin\psi^{(l,k)}}
    {4\pi\left\|\mathbf{q}^{(l,k)}-\mathbf{p}_n\right\|^3}.
\end{align}

\subsubsection{Equivalent channel spanning from the UEs to the RF chains at the BSs}
According to (\ref{System_Model_1}), (\ref{System_Model_2}) and (\ref{System_Model_4}), the equivalent channel spanning from UE-$k$ to the RF chain at BS-$l$, denoted as $h_l^{(k)}$, can be represented as
\begin{align}\label{Channel_Model_7}
    \notag h_l^{(k)}=&\mathbf{f}^{(l)\mathrm{H}}\mathbf{\Theta}^{(l)}\mathbf{g}^{(l,k)}\\
    =&\sum_{n=1}^N \sqrt{\varsigma_n^{(l)}\beta_n^{(l,k)}}\mathrm{e}^{\jmath
    \left(\overline{\theta}_n^{(l)}+\widetilde{\theta}_n^{(l)}
    -\frac{2\pi}{\lambda}\left(\left\|\mathbf{r}-\mathbf{p}_n\right\|
    +\left\|\mathbf{q}^{(l,k)}-\mathbf{p}_n\right\|\right)\right)}.
\end{align}
Note that the aggregated channel is contaminated by the RHS phase shift error. Given the random nature of phase shift error $\widetilde{\theta}_n^{(l)}$, only the mean of $h_l^{(k)}$ can be acquired by relying on the statistics of $\widetilde{\theta}_n^{(l)}$. Therefore, we have:
\begin{align}\label{Channel_Model_8}
    \overline{h}_l^{(k)}=\mathbb{E}[h_l^{(k)}],
\end{align}
and
\begin{align}\label{Channel_Model_9}
    \widetilde{h}_l^{(k)}=h_l^{(k)}-\mathbb{E}[h_l^{(k)}],
\end{align}
respectively.

\section{Beamforming Design}\label{Beamforming_Design}
Practical RF chain circuits generally suffer from HWIs, including power amplifier non-linearities, amplitude/phase imbalance in the In-phase/Quadrature mixers, phase error in the local oscillator, sampling jitter and finite-resolution quantization in the analog-to-digital converters~\cite{bjornson2017massive}. To characterize the impact of RF chain HWIs, the non-ideal hardware circuits of the transmitter and the receiver can be modelled as non-linear memoryless filters~\cite{bjornson2014massive}. Specifically, the key modeling characteristics in this non-linear memoryless filter are that the desired signal is scaled by a deterministic factor and that an uncorrelated memoryless signal distortion term is added, which follows the Gaussian distribution in the worst case. Upon considering the RF chains HWIs of the transceivers, the signal received by BS-$l$ is formulated as:
\begin{align}\label{Beamforming_Design_1}
    \notag y_l=&\sum_{k=1}^{K}\left(\sqrt{\rho_k\varepsilon_{u}\varepsilon_{v}}h_l^{(k)}s_k
    +\sqrt{\rho_k\left(1-\varepsilon_{u}\right)\varepsilon_{v}}h_l^{(k)}u^{(k)}\right.\\
    &\left.+\sqrt{\rho_k\left(1-\varepsilon_{v}\right)}h_l^{(k)}v_l^{(k)}\right)+w_l,
\end{align}
where $s_k$ is the desired information received from UE-$k$, $\rho_k$ denotes the transmit power of UE-$k$, and $w_l\sim\mathcal{CN}(0,\sigma_w^2)$ is the additive noise at the RF chain of BS-$l$. Furthermore, $u^{(k)}\sim\mathcal{CN}(0,1)$ and $v_l^{(k)}\sim\mathcal{CN}(0,1)$ are the distortion of the information symbol $s_k$ due to the RF chains HWIs of UE-$k$ and BS-$l$, respectively. Finally, $\varepsilon_{u}$ and $\varepsilon_{v}$ are the hardware quality factors of the RF chains at the UEs and the BSs, respectively, satisfying $0\leq\varepsilon_{u}\leq1$ and $0\leq\varepsilon_{v}\leq1$. Explicitly, a hardware quality factor of 1 indicates
that the hardware is ideal, while 0 means that the hardware is
completely inadequate. Based on the fronthaul between the CPU and the BSs, the CPU uses the signals $y_1,y_2,\cdots,y_L$ received from all BSs and designs the digital beamformer weights to recover the information from all UEs. Therefore, the signal received at the CPU is given by
\begin{align}\label{Beamforming_Design_2}
    \notag\mathbf{y}=&\left[y_1,y_2,\cdots,y_L\right]^\mathrm{T}\\
    \notag=&\sum_{k=1}^{K}\left(\sqrt{\rho_k\varepsilon_{u}\varepsilon_{v}}\mathbf{h}^{(k)}s_k
    +\sqrt{\rho_k\left(1-\varepsilon_{u}\right)\varepsilon_{v}}\mathbf{h}^{(k)}u^{(k)}\right.\\
    &\left.+\sqrt{\rho_k\left(1-\varepsilon_{v}\right)}\mathbf{h}^{(k)}\odot\mathbf{v}^{(k)}\right)
    +\mathbf{w}.
\end{align}

If we denote the receiver combining (RC) vector used for recovering the information $s_k$ as $\mathbf{b}^{(k)}$, the equivalent signal used for recovering the information $s_k$, denoted as $\mathbf{y}_k=\mathbf{b}^{(k)\mathrm{H}}\mathbf{y}$, is formulated in (\ref{Beamforming_Design_3}).
\begin{figure*}[!t]
\begin{align}\label{Beamforming_Design_3}
    \notag\mathbf{y}_k=&\mathbf{b}^{(k)\mathrm{H}}\left(\sum_{i=1}^{K}
    \left(\sqrt{\rho_i\varepsilon_{u}\varepsilon_{v}}\mathbf{h}^{(i)}s_i
    +\sqrt{\rho_i\left(1-\varepsilon_{u}\right)\varepsilon_{v}}\mathbf{h}^{(i)}u^{(i)}
    +\sqrt{\rho_i(1-\varepsilon_{v})}\mathbf{h}^{(i)}\odot\mathbf{v}\right)+\mathbf{w}\right)\\
    \notag=&\underbrace{\sqrt{\rho_k\varepsilon_{u}\varepsilon_{v}}
    \mathbf{b}^{(k)\mathrm{H}}\overline{\mathbf{h}}^{(k)}s_k}
    _{\text{Desired signal for $s_k$ over determinate channel}}
    +\underbrace{\sqrt{\rho_k\varepsilon_{u}\varepsilon_{v}}
    \mathbf{b}^{(k)\mathrm{H}}\widetilde{\mathbf{h}}^{(k)}s_k}
    _{\text{Desired signal for $s_k$ over unknown channel}}\\
    \notag&+\underbrace{\sqrt{\rho_k\left(1-\varepsilon_{u}\right)
    \varepsilon_{v}}\mathbf{b}^{(k)\mathrm{H}}
    \mathbf{h}^{(k)}u^{(k)}}_{\text{UE HWI distortion on $s_k$}}
    +\underbrace{\sqrt{\rho_k\left(1-\varepsilon_{v}\right)}\mathbf{b}^{(k)\mathrm{H}}
    \left(\mathbf{h}^{(k)}\odot\mathbf{v}\right)}_{\text{BS HWI distortion on $s_k$}}\\
    &+\underbrace{\mathbf{b}^{(k)\mathrm{H}}\mathop{\sum_{k'=1}^K}_{k'\neq k}
    \left(\sqrt{\rho_{k'}\varepsilon_{u}\varepsilon_{v}}\mathbf{h}^{(k')}s_{k'}
    +\sqrt{\rho_{k'}\left(1-\varepsilon_{u}\right)\varepsilon_{v}}\mathbf{h}^{(k')}u^{(k')}
    +\sqrt{\rho_{k'}\left(1-\varepsilon_{v}\right)}\mathbf{h}^{(k')}
    \odot\mathbf{v}\right)}_{\text{Inter-user interference on $s_k$}}
    +\underbrace{\mathbf{b}^{(k)\mathrm{H}}\mathbf{w}}_{\text{Additive noise on $s_k$}}
\end{align}
\hrulefill
\end{figure*}
Therefore, the instantaneous achievable rate of UE-$k$, denoted as $R_k$, is $R_k=\log_2(1+\gamma_k)$. Furthermore, $\gamma_k$ is the signal-to-interference-plus-noise ratio (SINR) of $s_k$ given in (\ref{Beamforming_Design_5}), where the correlation matrix
$\mathbf{C}_{\mathbf{h}^{(k)}\mathbf{h}^{(k)}}$ can be derived as shown in (\ref{Appendix_A_8}).
\begin{figure*}[!t]
\begin{align}\label{Beamforming_Design_5}
    \gamma_k=\frac{\rho_k\varepsilon_{u}\varepsilon_{v}
    \left|\mathbf{b}^{(k)\mathrm{H}}\overline{\mathbf{h}}^{(k)}\right|^2}{\mathbf{b}^{(k)\mathrm{H}}
    \left(\sum\limits_{k'=1}^K\rho_{k'}
    \left(\varepsilon_{v}\mathbf{C}_{\mathbf{h}^{(k')}\mathbf{h}^{(k')}}
    +\left(1-\varepsilon_{v}\right)\mathbf{C}_{\mathbf{h}^{(k')}\mathbf{h}^{(k')}}
    \odot\mathbf{I}_L\right)-\rho_k\varepsilon_{u}\varepsilon_{v}
    \overline{\mathbf{h}}^{(k)}\overline{\mathbf{h}}^{(k)\mathrm{H}}
    +\sigma_w^2\mathbf{I}_L\right)\mathbf{b}^{(k)}}
\end{align}
\hrulefill
\end{figure*}

For UE-$k$, we aim for jointly optimizing the active beamformer $\mathbf{b}^{(k)}$ at the CPU and the holographic beamformer $\overline{\mathbf{\Theta}}^{(1)},\overline{\mathbf{\Theta}}^{(2)},\cdots,
\overline{\mathbf{\Theta}}^{(L)}$ at all the distributed BSs for maximizing the instantaneous achievable rate $R_k$. The corresponding optimization problem can be formulated as
\begin{align}\label{Beamforming_Design_6}
    \text{(P1)}&\max_{\mathbf{b}^{(k)},
    \overline{\mathbf{\Theta}}^{(1)},\overline{\mathbf{\Theta}}^{(2)},\cdots,
    \overline{\mathbf{\Theta}}^{(L)}}R_k,\quad k=1,2,\cdots,K\\
    \text{s.t.}&\quad \overline{\mathbf{\Theta}}^{(l)}\overline{\mathbf{\Theta}}^{(l)\mathrm{H}}=\mathbf{I}_N,
    \quad l=1,2,\cdots,L.
\end{align}
To solve Problem (P1), it is expected that the CSI of all BSs is shared with all BSs, which requires sharing the CSI between all BSs. Hence, to reduce the overhead of sharing the CSI among all BSs, we employ hybrid beamforming. Specifically, the holographic beamformer at each distributed BS is designed based on the local CSI for maximizing the channel's power gain $\|\overline{\mathbf{h}}^{(1)}\|^2,\|\overline{\mathbf{h}}^{(2)}\|^2,\cdots,
\|\overline{\mathbf{h}}^{(K)}\|^2$, while the MMSE-based digital RC is employed at the CPU based on the overall CSI of the entire network for mitigating the inter-user interference and for recovering the information $s_1,s_2,\cdots,s_K$.

In the following, we present the details of the holographic beamformer harnessed at the BSs and the digital beamformer at the CPU.

\subsection{Holographic Beamforming}\label{Holographic_Beamforming}
For UE-$k$, we aim for jointly optimizing the holographic beamformer $\mathbf{\Theta}^{(1)},\mathbf{\Theta}^{(2)},\cdots,\mathbf{\Theta}^{(L)}$ in order to maximize the channel's power gain $\|\mathbf{h}^{(k)}\|^2$, formulated as:
\begin{align}\label{Holographic_Beamforming_1}
    \text{(P2)}&\max_{\overline{\mathbf{\Theta}}^{(l)},\ l\in\left\{1,2,\cdots,L\right\}}
    \left\|\ddot{\mathbf{h}}^{(k)}\right\|^2\\
    \text{s.t.}&\quad \overline{\mathbf{\Theta}}^{(l)}\overline{\mathbf{\Theta}}^{(l)\mathrm{H}}=\mathbf{I}_N,
    \quad l\in\left\{1,2,\cdots,L\right\},
\end{align}
where $\ddot{\mathbf{h}}^{(k)}$ is the equivalent channel in the absence of contamination from the RHS phase shift error, with the $l$th element given by
\begin{align}\label{Holographic_Beamforming_2}
    \ddot{h}_l^{(k)}=\sum\limits_{n=1}^N\sqrt{\varsigma_n^{(l)}\beta_n^{(l,k)}}\mathrm{e}^{\jmath
    \left(\overline{\theta}_n^{(l)}-\frac{2\pi}{\lambda}
    \left(\left\|\mathbf{r}-\mathbf{p}_n\right\|
    +\left\|\mathbf{q}^{(l,k)}-\mathbf{p}_n\right\|\right)\right)}.
\end{align}
This shows that since all BSs jointly support all $K$ UEs, the holographic beamformer of all BSs focused on a specific UE-$k$ for maximizing the channel's power gain $\|\ddot{\mathbf{h}}^{(k)}\|^2$ is at the cost of disregarding other $K-1$ UEs. Since the main advantage of the cell-free architecture is that of reducing the path-loss from each BS to its nearest UE, the holographic beamformer at a specific BS can be focused on its nearest UE. Specifically, we denote the BS set having its holographic beamformer focused on UE-$k$ as $\mathcal{L}_k=\{l:D_k^{(l)}\leq D_k^{(l')},l'=1,2,\cdots,L\}$ with $D_k^{(l)}$ being the distance from UE-$k$ to BS-$l$. Therefore, Problem (P2) can be reformulated as
\begin{align}\label{Holographic_Beamforming_3}
    \text{(P3)}&\max_{\overline{\mathbf{\Theta}}^{(l)},\ l\in\mathcal{L}_k}\left\|\ddot{\mathbf{h}}_{\mathcal{L}_k}^{(k)}\right\|^2\\
    \text{s.t.}&\quad \overline{\mathbf{\Theta}}^{(l)}\overline{\mathbf{\Theta}}^{(l)\mathrm{H}}=\mathbf{I}_N,
    \quad l\in\mathcal{L}_k,
\end{align}
where $\|\ddot{\mathbf{h}}_{\mathcal{L}_k}^{(k)}\|^2$ is the channel's power gain for UE-$k$ corresponding to the BSs belonging to the set $\mathcal{L}_k$, given by $\|\ddot{\mathbf{h}}_{\mathcal{L}_k}^{(k)}\|^2
=\sum_{l\in\mathcal{L}_k}|\ddot{h}_l^{(k)}|^2$. Therefore, the holographic beamformer in BS-$l$ of Problem (P3) is optimized as
\begin{align}\label{Holographic_Beamforming_5}
    \overline{\theta}_n^{(l)}=\frac{2\pi}{\lambda}\left(\left\|\mathbf{r}-\mathbf{p}_n\right\|
    +\left\|\mathbf{q}^{(l,k)}-\mathbf{p}_n\right\|\right),\ l\in\mathcal{L}_k.
\end{align}
Upon substituting (\ref{Holographic_Beamforming_5}) into (\ref{Holographic_Beamforming_2}), we arrive at
\begin{align}\label{Holographic_Beamforming_6}
    \ddot{h}_l^{(k)}=\sum_{n=1}^N\sqrt{\varsigma_n\beta_n^{(l,k)}}
\end{align}
if $l\in\mathcal{L}_k$, or at
\begin{align}\label{Holographic_Beamforming_7}
    \ddot{h}_l^{(k)}=\sum_{n=1}^N \sqrt{\varsigma_n\beta_n^{(l,k)}}\mathrm{e}^{\jmath
    \frac{2\pi}{\lambda}\left(\left\|\mathbf{q}^{(l,k')}-\mathbf{p}_n\right\|
    -\left\|\mathbf{q}^{(l,k)}-\mathbf{p}_n\right\|\right)}
\end{align}
if $l\in\mathcal{L}_{k'}$ with $k'\neq k$.

\subsection{Digital Beamforming}\label{Digital_Beamforming}
Upon getting the optimized holographic beamformer matrices $\overline{\mathbf{\Theta}}^{(1)},\overline{\mathbf{\Theta}}^{(2)},\cdots,
\overline{\mathbf{\Theta}}^{(L)}$, the equivalent channels spanning from all $K$ UEs to the CPU, i.e $\ddot{\mathbf{h}}^{(1)},\ddot{\mathbf{h}}^{(2)},\cdots,\ddot{\mathbf{h}}^{(K)}$, can be obtained. Then, the optimal active beamformer $\mathbf{b}^{(k)}$ for the recovery of the information $s_k$ can be designed based on the MMSE criterion as
\begin{align}
    \notag\mathbf{b}^{(k)}=&\rho_{k}\varepsilon_u\varepsilon_v\left(\sum_{k'=1}^{K}\rho_{k'}
    \left(\varepsilon_v\mathbf{C}_{{\mathbf{h}}^{(k')}{\mathbf{h}}^{(k')}}\right.\right.\\
    &\left.\left.+\left(1-\varepsilon_v\right)\mathbf{C}_{{\mathbf{h}}^{(k')}
    {\mathbf{h}}^{(k')}}\odot\mathbf{I}_{L}\right)+\sigma_w^2\mathbf{I}_L\right)^\mathrm{-1}
    \ddot{\mathbf{h}}^{(k)}.
\end{align}

\section{Theoretical Analysis}\label{Theoretical_Analysis}
In this section, we theoretically derive the ergodic achievable sum-rate upper bound, given the random distribution of BSs following a homogeneous PPP, by employing the popular stochastic geometry approach.

\begin{lemma}\label{Lemma_1}
     Since the distribution of the BSs follows the PPP, each realization of the BS generation, including the number of BSs and their positions, are random. The instantaneous spectral efficiency of UE-$k$ with respect to the channel $\ddot{\mathbf{h}}^{(k)}$ in each realization of BS generation is given by
    \begin{align}\label{Theoretical_Analysis_1}
        R_k=\log_2\left(1+\gamma_k\right),
    \end{align}
    with the SINR $\gamma_k$ formulated as:
    \begin{align}\label{Theoretical_Analysis_2}
        \notag\gamma_k=&\rho_k\varepsilon_u\varepsilon_v\xi^2\ddot{\mathbf{h}}^{(k)\mathrm{H}}
        \left(\sum_{k'=1}^{K}\rho_{k'}\left(\varepsilon_v
        \xi^2\ddot{\mathbf{h}}^{(k')}\ddot{\mathbf{h}}^{(k')\mathrm{H}}\right.\right.\\
        \notag&\left.\left.+\left(1-\varepsilon_v\right)\xi^2\left(\ddot{\mathbf{h}}^{(k')}
        \ddot{\mathbf{h}}^{(k')\mathrm{H}}\right)
        \odot\mathbf{I}_{L}+\left(1-\xi^2\right)\mathbf{Q}^{(k')}\right)\right.\\
        &\left.-\rho_k\varepsilon_u\varepsilon_v\xi^2\ddot{\mathbf{h}}^{(k)}
        \ddot{\mathbf{h}}^{(k)\mathrm{H}}
        +\sigma_w^2\mathbf{I}_{L}\right)^{-1}\ddot{\mathbf{h}}^{(k)}.
    \end{align}
    In (\ref{Theoretical_Analysis_2}) we have
    \begin{align}
        \mathbf{Q}^{(k)}=\mathbf{Diag}
        \left\{\left\|\nu_{1}^{(k)}\right\|^2,\left\|\nu_{2}^{(k)}\right\|^2,\cdots,
        \left\|\nu_{L}^{(k)}\right\|^2\right\}
    \end{align}
    with $\nu_l^{(k)}=\left[\sqrt{\varsigma_1^{(l)}\beta_1^{(l,k)}},\sqrt{\varsigma_2^{(l)}
    \beta_2^{(l,k)}},\cdots,\sqrt{\varsigma_N^{(l)}\beta_N^{(l,k)}}\right]^\mathrm{T}$.
    Furthermore, $\xi=\frac{\sin(\iota_\mathrm{p})}{\iota_\mathrm{p}}$ when the RHS phase error follows $\mathcal{U}(-\iota_\mathrm{p},\iota_\mathrm{p})$, and $\xi=\frac{I_1(\varpi_\text{p})}{I_0(\varpi_\mathrm{p})}$ when it obeys $\mathcal{VM}(0,\varpi_\text{p})$ with $I_0(\cdot)$ and $I_1(\cdot)$ representing the modified Bessel functions of the first kind of order 0 and order 1, respectively. The RHS phase error power is $\sigma_\mathrm{p}^2=\mathbb{E}[\widetilde{\theta}_n^{(l)2}]=\frac{1}{3}\iota_\mathrm{p}^2$ and $\sigma_\mathrm{p}^2=\mathbb{E}[\widetilde{\theta}_n^{(l)2}]=\frac{1}{\varpi_\text{p}}$, when it follows the uniform distribution and the von-Mises distribution, respectively.
\end{lemma}
\begin{IEEEproof}
    See Appendix~\ref{Appendix_A}.
\end{IEEEproof}

According to (\ref{Theoretical_Analysis_1}), the ergodic sum-rate can be expressed as $R=\mathbb{E}[\sum_{k=1}^KR_k]$.

\begin{theorem}\label{Theorem_1}
The ergodic sum-rate upper bound, denoted as $\overline{R}$, can be represented as shown in (\ref{Theoretical_Analysis_4}), where we have $\boldsymbol{\varsigma}\in\mathbb{C}^{N\times1}$ with $\varsigma_n$ given in (\ref{System_Model_3}), and $\boldsymbol{\beta}^{(o)}\in\mathbb{C}^{N\times1}$ with $\beta_n^{(o)}=\frac{1}{4\pi^2}\iint_{\mathbf{c}\in\mathcal{C}_o}\int_{0}^{\pi}
\frac{A\|\mathbf{c}\|\sin\omega}{(\|\mathbf{c}\|^2+2\|\mathbf{c}\|x_n\cos\omega+x_n^2
+(H+y_n)^2)^\frac{3}{2}}\mathrm{d}\omega\mathrm{d}\mathbf{c}$. Furthermore, $S$ represents the area of the BS distribution and $\mathcal{C}_o$ is the BS distribution area with the geometric center as the origin of the coordinate system.
\begin{figure*}[!t]
\begin{align}\label{Theoretical_Analysis_4}
    \overline{R}=\sum_{k=1}^K\log_2\left(1+\frac{\rho_k\varepsilon_u\varepsilon_v\xi^2
    \frac{\eta}{K}\cdot\left|\sqrt{\boldsymbol{\varsigma}}^\mathrm{H}
    \sqrt{\boldsymbol{\beta}^{(o)}}\right|^2}{\rho_k\left(\frac{1}{\eta S}
    \left(1-\varepsilon_v\right)+\frac{1}{K}\left(1-\varepsilon_u\right)\varepsilon_v\right)\xi^2
    \eta\cdot\left|\sqrt{\boldsymbol{\varsigma}}^\mathrm{H}\sqrt{\boldsymbol{\beta}^{(o)}}\right|^2
    +\rho_k\frac{1}{S}\left(1-\xi^2\right)\cdot\boldsymbol{\varsigma}^\mathrm{H}
    \boldsymbol{\beta}^{(o)}+\sigma_w^2}\right)
\end{align}
\hrulefill
\end{figure*}
\end{theorem}
\begin{IEEEproof}
    See Appendix~\ref{Appendix_B}.
\end{IEEEproof}

\begin{corollary}\label{Corollary_1}
When the transmit power $\rho_k\rightarrow\infty$, the ergodic achievable rate upper bound is given in (\ref{Theoretical_Analysis_11}). This indicates that the ergodic achievable sum-rate is limited by the hardware quality of the RHS elements as well as of the RF chains at the UEs and the BSs. Hence there is a saturation in the high transmit power region, when the hardware quality is non-ideal.
\begin{figure*}[!t]
\begin{align}\label{Theoretical_Analysis_11}
    \overline{R}=K\log_2\left(1+\frac{\varepsilon_u\varepsilon_v\xi^2\frac{\eta}{K}\cdot
    \left|\sqrt{\boldsymbol{\varsigma}}^\mathrm{H}\sqrt{\boldsymbol{\beta}^{(o)}}\right|^2}
    {\left(\frac{1}{\eta S}\left(1-\varepsilon_v\right)
    +\frac{1}{K}\left(1-\varepsilon_u\right)\varepsilon_v\right)\xi^2\eta\cdot
    \left|\sqrt{\boldsymbol{\varsigma}}^\mathrm{H}\sqrt{\boldsymbol{\beta}^{(o)}}\right|^2
    +\frac{1}{S}\left(1-\xi^2\right)\cdot
    \boldsymbol{\varsigma}^\mathrm{H}\boldsymbol{\beta}^{(o)}}\right)
\end{align}
\hrulefill
\end{figure*}
\end{corollary}
\begin{IEEEproof}
    It can be directly obtained by setting $\rho\rightarrow\infty$ in (\ref{Theoretical_Analysis_4}).
\end{IEEEproof}

\begin{corollary}\label{Corollary_2}
We analyze the impact of hardware quality of the RHS elements, of the RF chains at the UEs and of the RF chains at the BSs on the ergodic achievable sum-rate upper bound as follows.

Firstly, when the hardware quality of the RHS elements and the RF chains at the BSs is ideal, while that of the RF chains at the UEs is non-ideal, i.e. we have $\xi=1$, $\varepsilon_v=1$ and $\varepsilon_u<1$, the ergodic achievable sum-rate upper bound becomes
\begin{align}\label{Theoretical_Analysis_11_1}
    \overline{R}=\sum_{k=1}^K\log_2\left(1+\frac{\rho_k\varepsilon_u\frac{\eta}{K}\cdot
    \left|\sqrt{\boldsymbol{\varsigma}}^\mathrm{H}\sqrt{\boldsymbol{\beta}^{(o)}}\right|^2}
    {\rho_k\left(1-\varepsilon_u\right)\frac{\eta}{K}\cdot
    \left|\sqrt{\boldsymbol{\varsigma}}^\mathrm{H}\sqrt{\boldsymbol{\beta}^{(o)}}\right|^2
    +\sigma_w^2}\right).
\end{align}

Secondly, when the hardware quality of the RHS elements and the RF chains of the UEs are ideal, while that of the RF chains at the BSs is non-ideal, i.e. we have $\xi=1$, $\varepsilon_u=1$ and $\varepsilon_v<1$, the ergodic achievable sum-rate upper bound becomes
\begin{align}\label{Theoretical_Analysis_11_2}
    \overline{R}=\sum_{k=1}^K\log_2\left(1+\frac{\rho_k\varepsilon_v\frac{\eta}{K}\cdot
    \left|\sqrt{\boldsymbol{\varsigma}}^\mathrm{H}\sqrt{\boldsymbol{\beta}^{(o)}}\right|^2}
    {\rho_k\frac{1}{S}\left(1-\varepsilon_v\right)
    \left|\sqrt{\boldsymbol{\varsigma}}^\mathrm{H}\sqrt{\boldsymbol{\beta}^{(o)}}\right|^2
    +\sigma_w^2}\right).
\end{align}

Finally, when the hardware quality of the RF chains at both the UEs and BSs is ideal, while that of the RHS elements is non-ideal, i.e. we have $\varepsilon_u=1$, $\varepsilon_v=1$ and $\xi<1$, the ergodic achievable sum-rate upper bound becomes
\begin{align}\label{Theoretical_Analysis_11_3}
    \overline{R}=\sum_{k=1}^K\log_2\left(1+\frac{\rho_k\xi^2\frac{\eta}{K}\cdot
    \left|\sqrt{\boldsymbol{\varsigma}}^\mathrm{H}\sqrt{\boldsymbol{\beta}^{(o)}}\right|^2}
    {\rho_k\frac{1}{S}\left(1-\xi^2\right)\cdot\boldsymbol{\varsigma}^\mathrm{H}
    \boldsymbol{\beta}^{(o)}+\sigma_w^2}\right).
\end{align}

In the high transmit power region, i.e. $\rho_k\rightarrow\infty$, the asymptotic ergodic sum-rate upper bound of (\ref{Theoretical_Analysis_11_1}), (\ref{Theoretical_Analysis_11_2}) and (\ref{Theoretical_Analysis_11_3}) tends to $\overline{R}=K\log_2\left(1+\frac{\varepsilon_u}{1-\varepsilon_u}\right)$, $\overline{R}=K\log_2\left(1+\frac{\varepsilon_v\eta S}{K(1-\varepsilon_v)}\right)$ and $\overline{R}=K\log_2\left(1+\frac{\xi^2\eta S\cdot|\sqrt{\boldsymbol{\varsigma}}^\mathrm{H}
\sqrt{\boldsymbol{\beta}^{(o)}}|^2}{K(1-\xi^2)\cdot\boldsymbol{\varsigma}^\mathrm{H}
\boldsymbol{\beta}^{(o)}}\right)$, respectively. This means that increasing the BS density is capable of compensating the HWI of the RF chains at the BSs and the PSE at the RHS, but it cannot compensate for the HWI of the RF chains at the UEs.
\end{corollary}
\begin{IEEEproof}
    They can be directly obtained by setting $\xi=1$, $\varepsilon_v=1$ and $\varepsilon_u=1$ in (\ref{Theoretical_Analysis_4}), respectively.
\end{IEEEproof}

\begin{corollary}\label{Corollary_3}
When the BS density obeys $\eta\rightarrow\infty$, the ergodic achievable sum-rate upper bound $\overline{R}$ becomes $\overline{R}=K\log_2\left(1+\frac{\varepsilon_u}{1-\varepsilon_u}\right)$, which is the same as that of the ideal BS hardware quality and RHS elements. It further demonstrates that increasing the density of the BSs can compensate for the phase shift error at the RHS elements and for the HWIs of the RF chains at the BSs. But it cannot compensate for the HWIs of the RF chains at the UEs.
\end{corollary}
\begin{IEEEproof}
    It can be directly obtained by setting $\eta\rightarrow\infty$ in (\ref{Theoretical_Analysis_4}).
\end{IEEEproof}

\begin{theorem}\label{Theorem_2}
    In the case of an infinitely large physical size, i.e. $N_x\rightarrow\infty$ and $N_y\rightarrow\infty$, the ergodic achievable rate upper bound is given in (\ref{Theoretical_Analysis_14}),
    \begin{figure*}[!t]
        \begin{align}\label{Theoretical_Analysis_14}
            \overline{R}=\sum_{k=1}^K\log_2\left(1+
            \frac{\rho_k\varepsilon_u\varepsilon_v\xi^2\eta\zeta\epsilon^2/K}
            {\rho_k\left(\frac{1}{\eta S}\left(1-\varepsilon_v\right)
            +\frac{1}{K}\left(1-\varepsilon_u\right)\varepsilon_v\right)\xi^2\eta\zeta\epsilon^2
            +\rho_k\frac{1}{S}\left(1-\xi^2\right)\zeta+\sigma_w^2}\right)
        \end{align}
    \hrulefill
    \end{figure*}
    where $\zeta=\frac{1}{2\pi^2}\iint_{\mathbf{c}\in\mathcal{C}_o}
    \frac{A\|\mathbf{c}\|}{(\|\mathbf{c}\|^2+H^2)^\frac{3}{2}}\mathrm{d}\mathbf{c}$ and $\epsilon=\int_{-\infty}^{\infty}\int_{-\infty}^{\infty}\sqrt{\frac{(\alpha+1)
    d_0^{\alpha+1}}{2\pi A(d_0^2+x^2+y^2)^{\frac{\alpha+3}{2}}}}\mathrm{d}x\mathrm{d}y$.
\end{theorem}
\begin{IEEEproof}
    See Appendix~\ref{Appendix_C}.
\end{IEEEproof}

\section{Numerical and Simulation Results}\label{Numerical_and_Simulation_Results}
In this section, the average achievable rate of the near-field RHS architecture of cell-free networks is presented. Unless otherwise specified, the simulation parameters are: the wavelength is $\lambda=10^{-2}\mathrm{m}$, the BS density is $\eta=10^{-3}/\mathrm{m}^2$, the number of RHS elements at each BS is $N=64\times64$, the transmit power of all UEs are identical, i.e. $\rho=\rho_1=\rho_2=\cdots=\rho_K$, the additive noise power at the RF chains of the BSs is $\sigma_w^2=-120\mathrm{dB}$, the height of the BSs is $H=10\mathrm{m}$, the distance between the RF chain and the RHS is $d_0=0.2\mathrm{m}$, the physical size of each RHS element is $A=5\mathrm{mm}\times5\mathrm{mm}$, the gain of RF chains is $2(\alpha+1)=10\mathrm{dB}$, the hardware quality is $\varepsilon_u=\varepsilon_v=1$, and the RHS phase shift error power is $\sigma_\mathrm{p}^2=0$. We assume that the BSs are distributed over a disk having a radius of $100\mathrm{m}$ following a homogeneous PPP.

\begin{figure}[!t]
    \centering
    \includegraphics[width=2.7in]{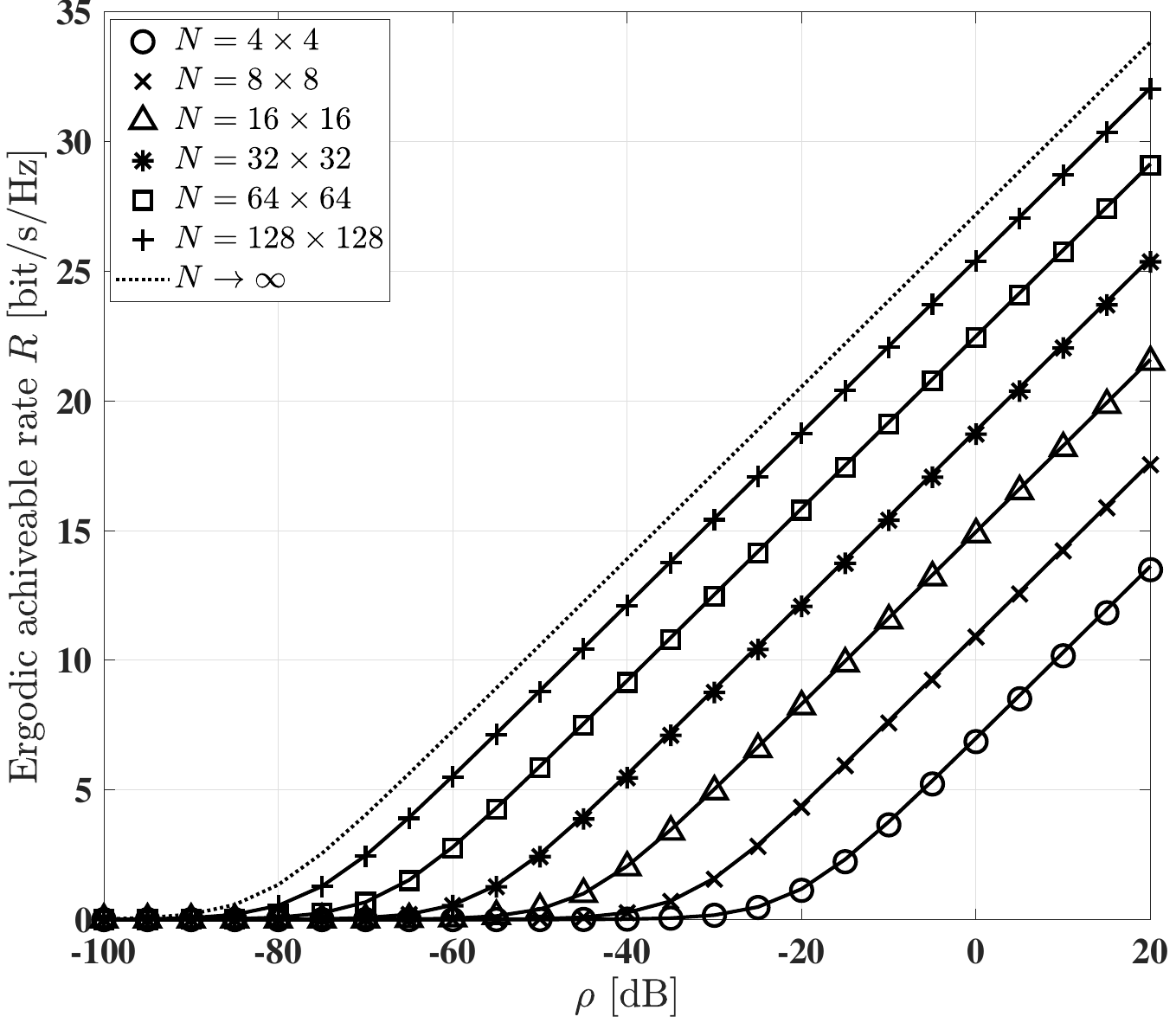}
    \caption{Theoretical analysis (\ref{Theoretical_Analysis_4}), (\ref{Theoretical_Analysis_14}) and simulation results of the ergodic achievable rate $R$ versus the transmit power $\rho$ for different number of RHS elements, with perfect RHS phase shift design and ideal hardware quality of the RF chains at the BSs and UEs.}\label{Fig_simu_cell_free_near_1}
\end{figure}

\begin{figure}[!t]
    \centering
    \subfloat[$\varepsilon_u=1$ and $\varepsilon_v<1$]
    {\begin{minipage}{1\linewidth}
        \centering
        \includegraphics[width=2.7in]{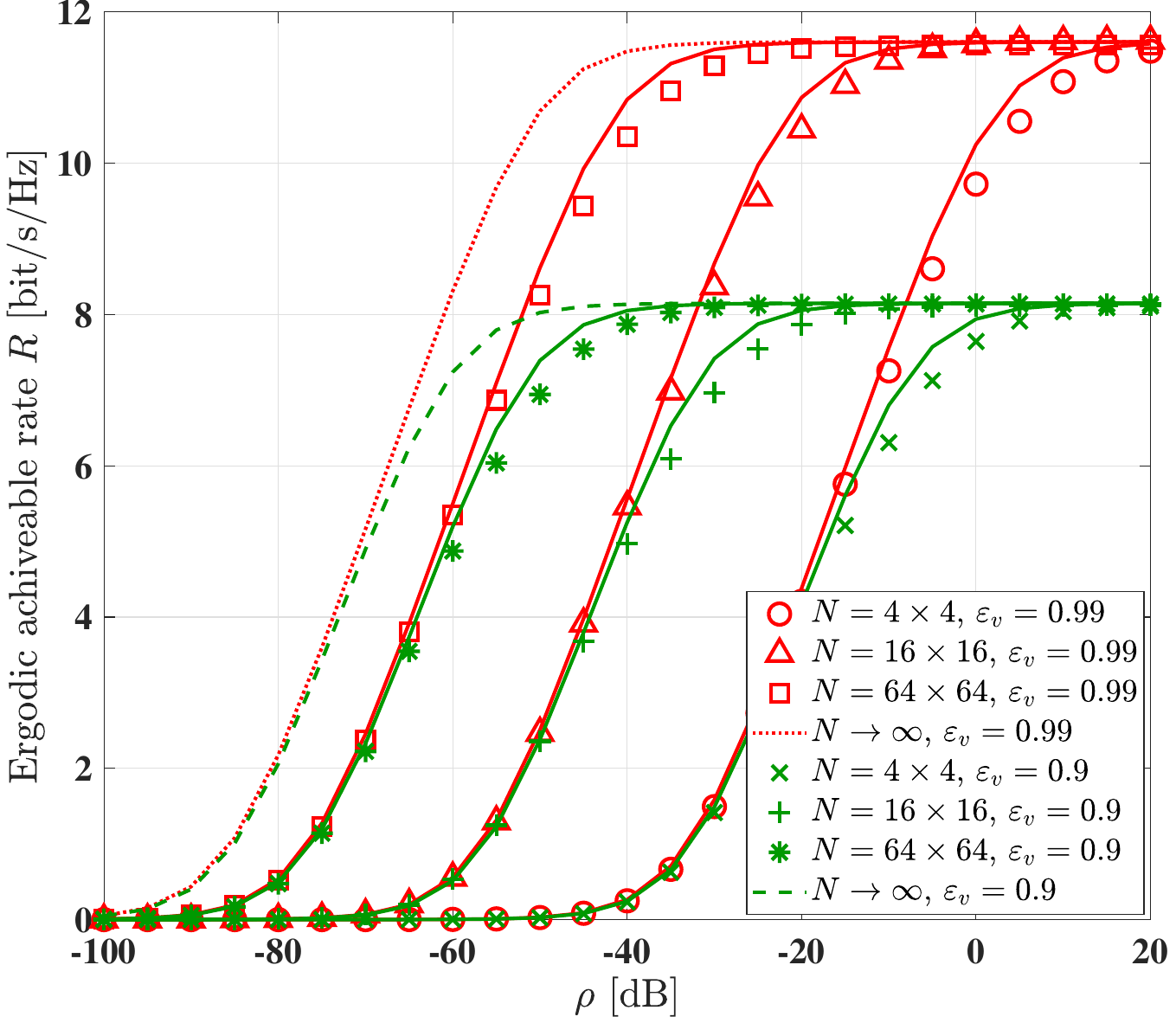}
    \end{minipage}}\\
    \subfloat[$\varepsilon_v=1$ and $\varepsilon_u<1$]
    {\begin{minipage}{1\linewidth}
        \centering
        \includegraphics[width=2.7in]{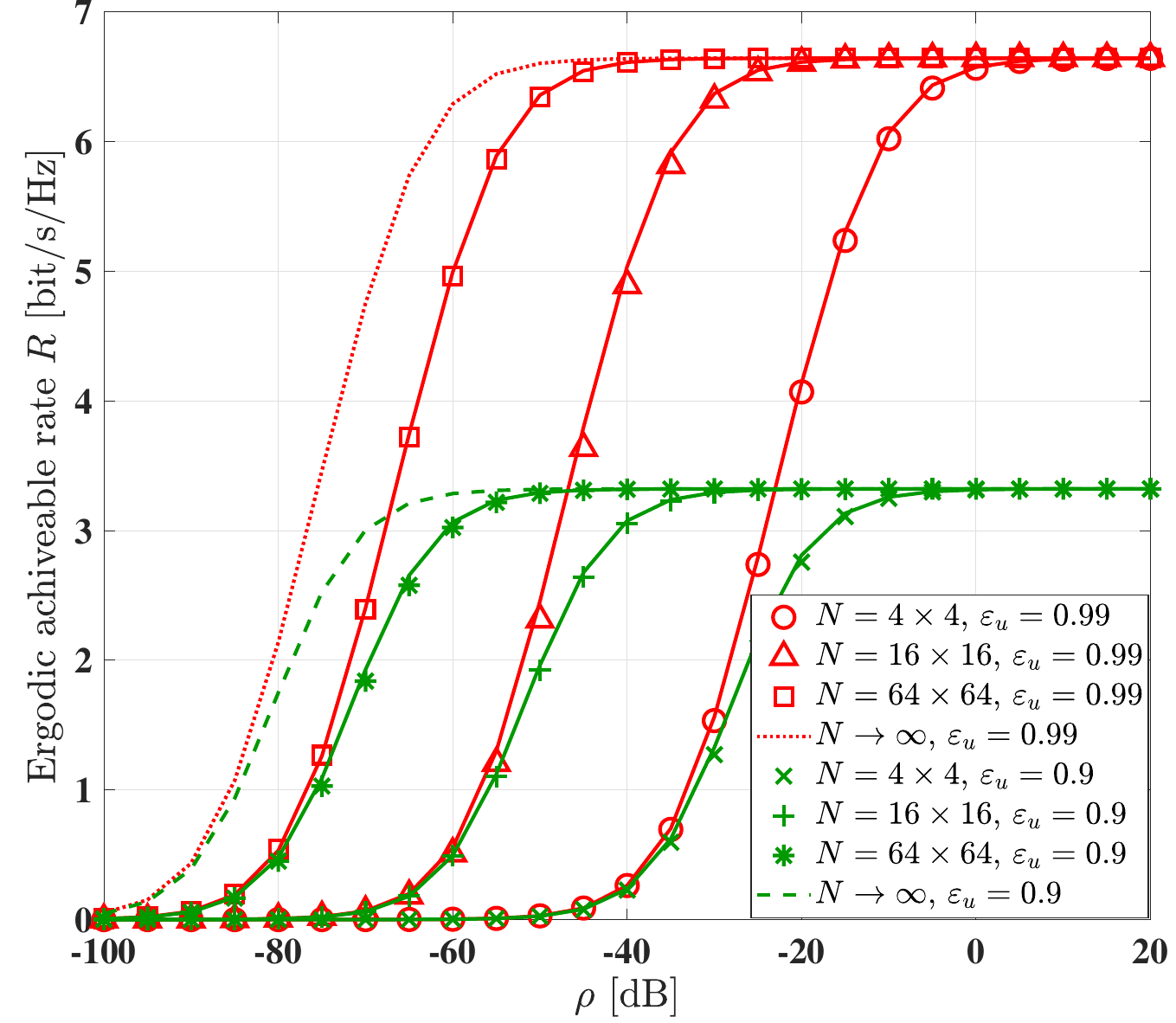}
    \end{minipage}}
    \caption{Theoretical analysis (\ref{Theoretical_Analysis_11_1}), (\ref{Theoretical_Analysis_11_2}), (\ref{Theoretical_Analysis_14}) and simulation results of the ergodic achievable rate $R$ versus the transmit power $\rho$ for different number of RHS elements.}\label{Fig_simu_cell_free_near_23}
\end{figure}

\begin{figure}[!t]
    \centering
    \subfloat[$\sigma_\mathrm{p}^2=0.1$]
    {\begin{minipage}{1\linewidth}
        \centering
        \includegraphics[width=2.7in]{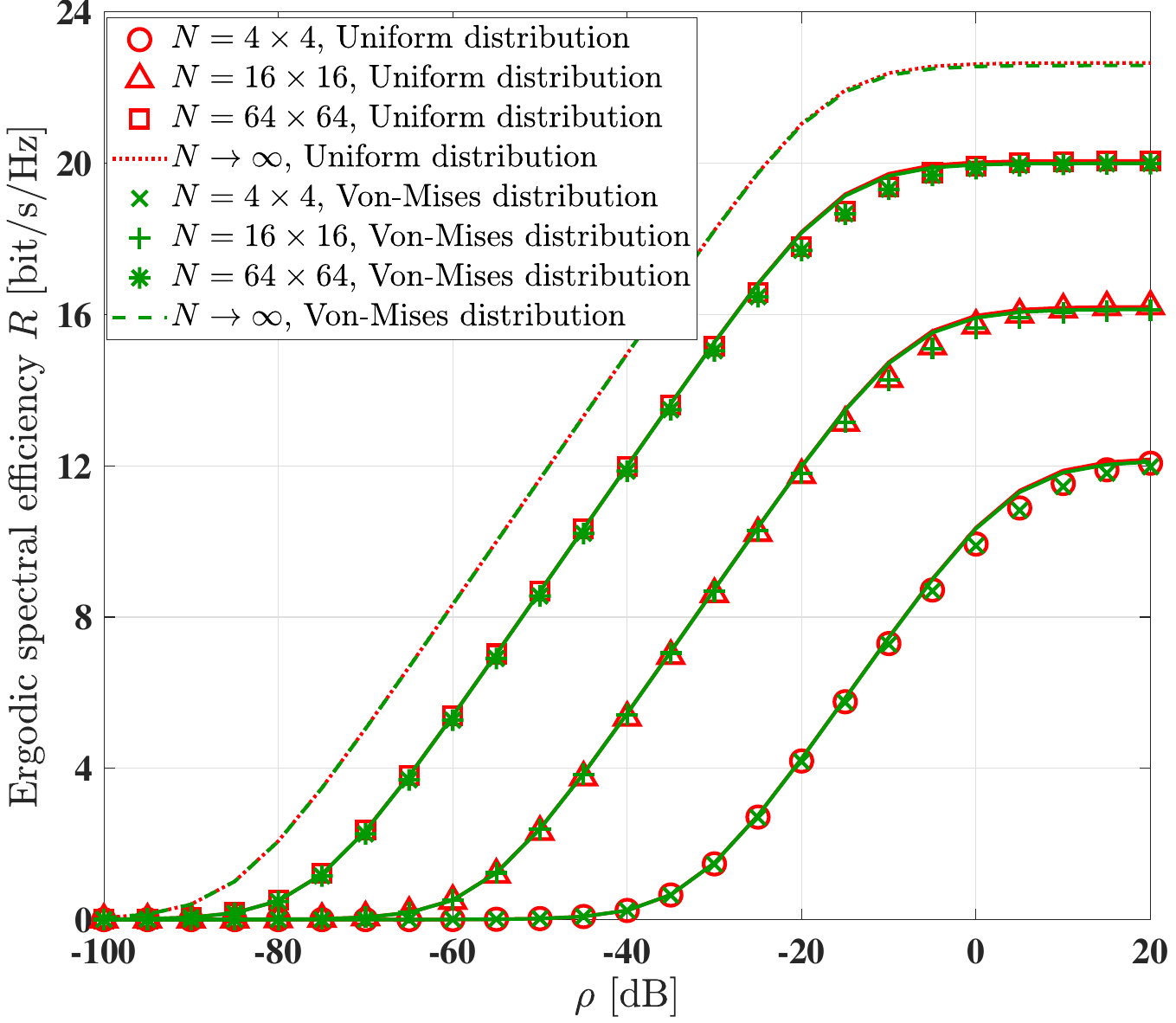}
    \end{minipage}}\\
    \subfloat[$\sigma_\mathrm{p}^2=1$]
    {\begin{minipage}{1\linewidth}
        \centering
        \includegraphics[width=2.7in]{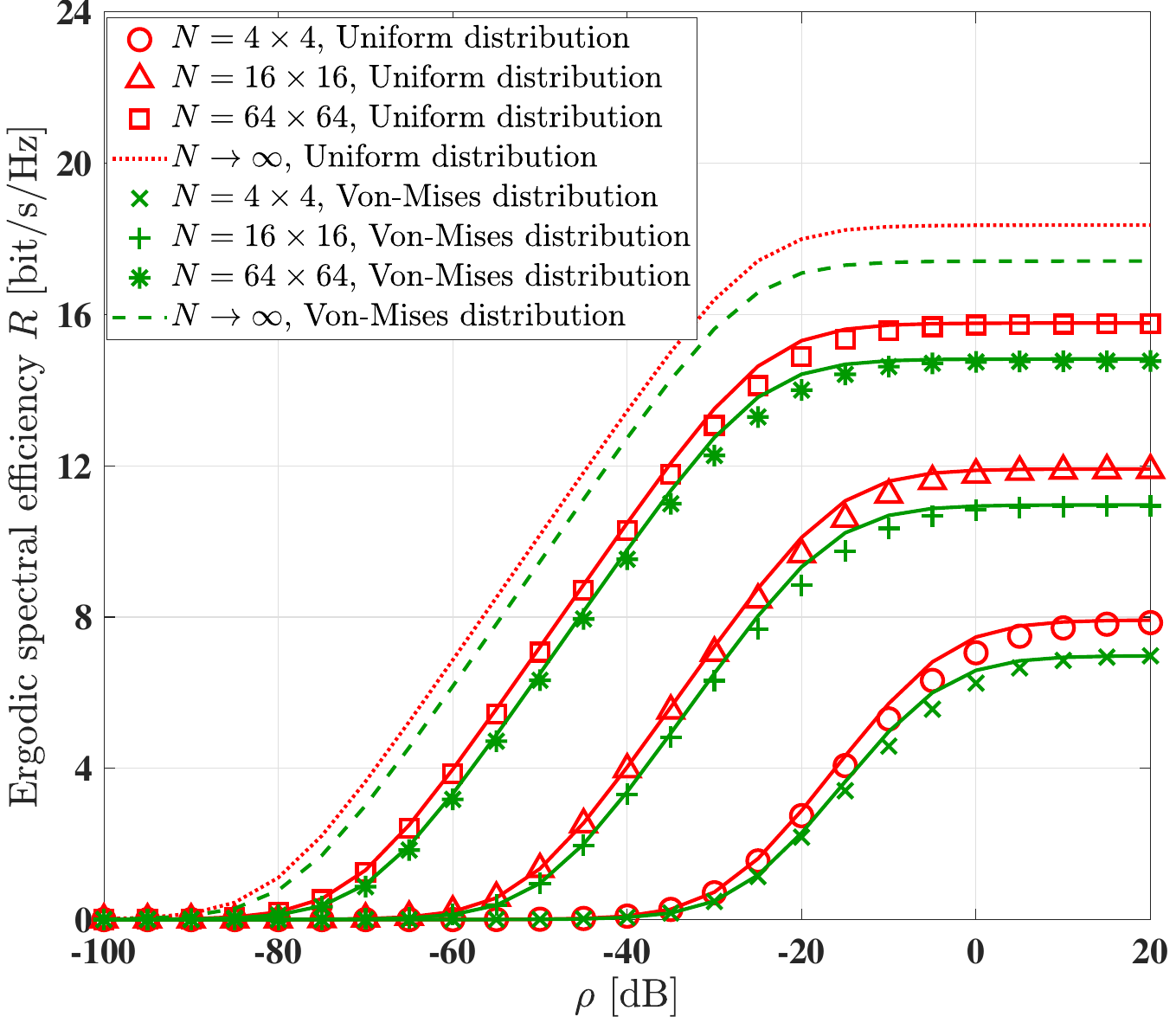}
    \end{minipage}}
    \caption{Theoretical analysis (\ref{Theoretical_Analysis_11_3}), (\ref{Theoretical_Analysis_14}) and simulation results of the ergodic achievable rate $R$ versus the transmit power $\rho$ for different number of RHS elements, with imperfect RHS phase shift design, i.e. $\sigma_\mathrm{p}^2>0$.}\label{Fig_simu_cell_free_near_45}
\end{figure}

Firstly, we focus our attention on the single-UE case, where the UE is located on the origin. In Fig.~\ref{Fig_simu_cell_free_near_1}, Fig.~\ref{Fig_simu_cell_free_near_23} and Fig.~\ref{Fig_simu_cell_free_near_45}, we utilize lines, e.g. `$--$', to represent the theoretical upper bound of the ergodic achievable rate, and utilize markers, e.g. `$\square$', to represent the corresponding simulation results.

Fig.~\ref{Fig_simu_cell_free_near_1} compares the ergodic achievable rate $R$ versus the transmit power $\rho$ for different number of RHS elements, for perfect RHS phase shift and ideal hardware quality of the RF chains at the BSs and UEs. As shown in Fig.~\ref{Fig_simu_cell_free_near_1}, the ergodic achievable rate is improved upon increasing the number of RHS elements. Furthermore, there is an achievable rate performance bound, when the number of RHS elements obeys $N\rightarrow\infty$, which is different from the far-field model assumption, where the achievable rate can increase infinitely. This is due to the factor that the far-field model assumption ignores the increased path-loss between the RF chain and the RHS elements, when $N\rightarrow\infty$.

Fig.~\ref{Fig_simu_cell_free_near_23} investigates the effect of HWI of the RF chains at the BSs and the UEs on the ergodic achievable rate. As shown in Fig.~\ref{Fig_simu_cell_free_near_23}, the achievable rate is limited by the hardware quality and it saturates in the high transmit power region. Furthermore, it shows that the realistic HWIs cannot be compensated by harnessing more RHS elements.

Fig.~\ref{Fig_simu_cell_free_near_45} presents the effect of the PSE of the RHS elements on the ergodic achievable rate, with the phase shift error power of $\sigma_\mathrm{p}^2=0.1$ and  $\sigma_\mathrm{p}^2=1$. Similar to the effect of the HWIs of the RF chains at the BSs and UEs, the achievable rate cannot be improved without limit in the high transmit power region. Furthermore, Fig.~\ref{Fig_simu_cell_free_near_45} shows that the ergodic achievable rate degrades upon increasing the phase shift error power $\sigma_\mathrm{p}^2$.

\begin{figure}[!t]
    \centering
    \subfloat[$\varepsilon_u=1$ and $\varepsilon_v=0.99$]
    {\begin{minipage}{1\linewidth}
        \centering
        \includegraphics[width=2.7in]{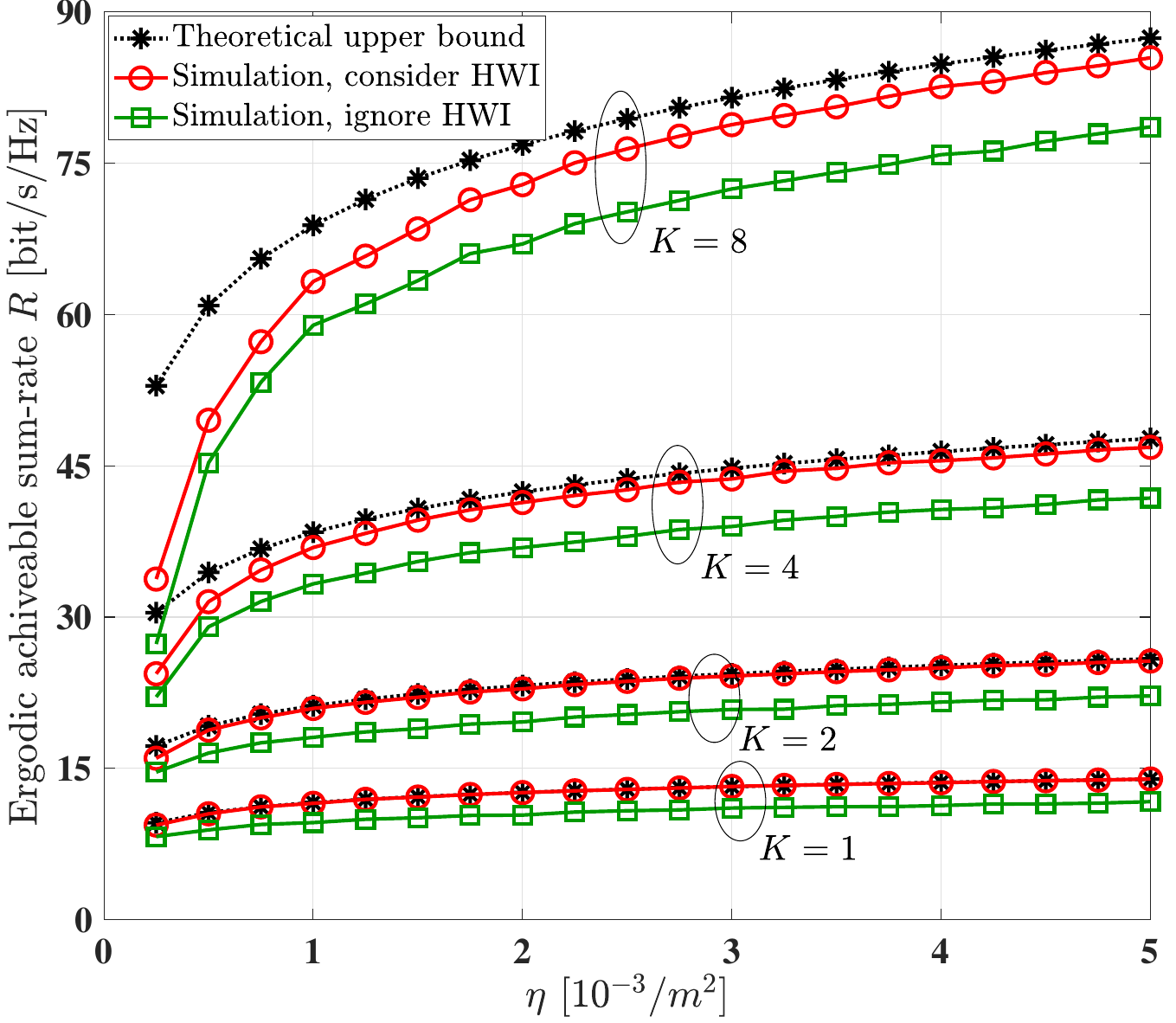}
    \end{minipage}}\\
    \subfloat[$\varepsilon_v=1$ and $\varepsilon_u=0.99$]
    {\begin{minipage}{1\linewidth}
        \centering
        \includegraphics[width=2.7in]{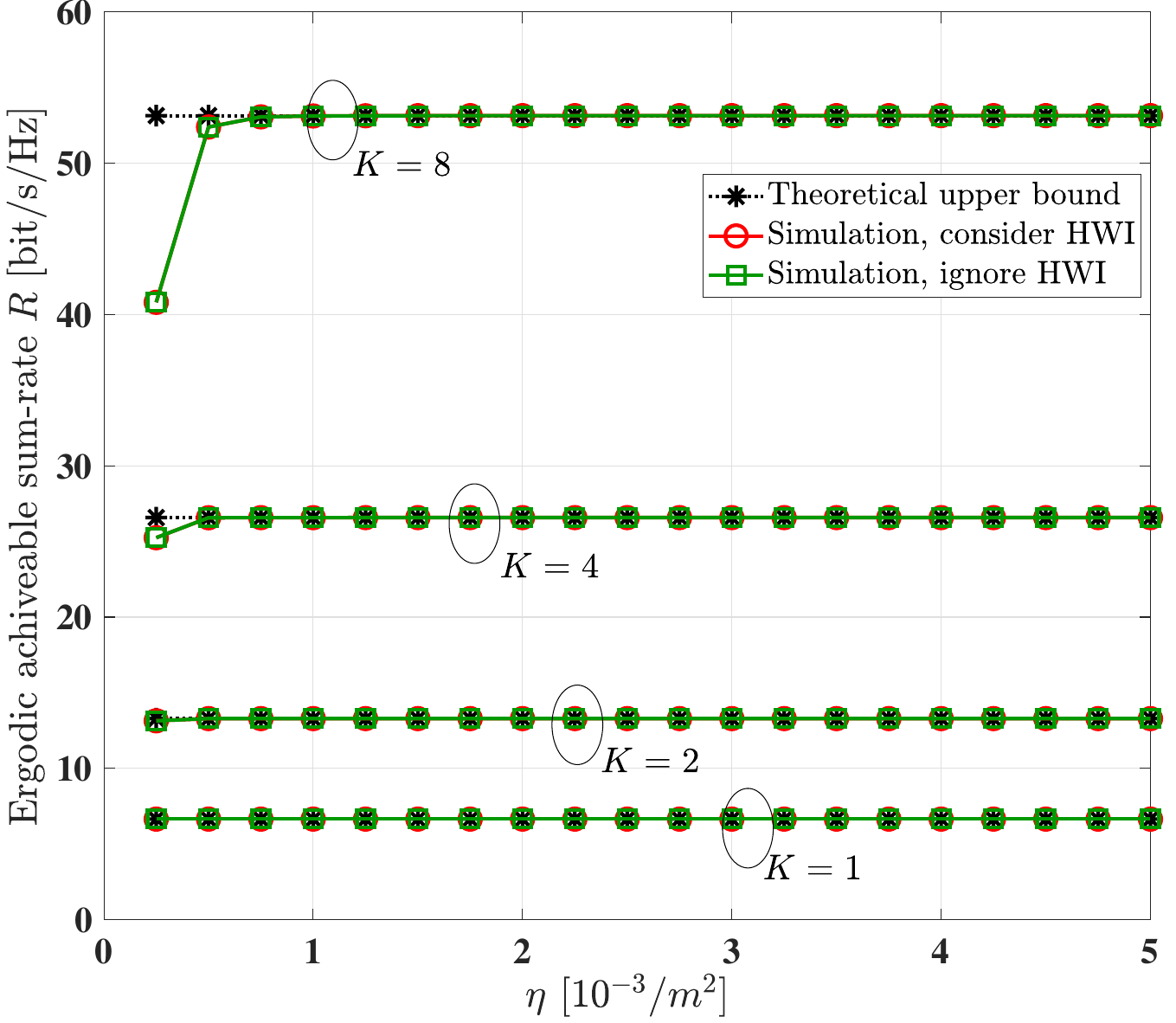}
    \end{minipage}}
    \caption{Theoretical analysis (\ref{Theoretical_Analysis_11_1}), (\ref{Theoretical_Analysis_11_2}) and simulation results of the achievable sum-rate $R$ versus the deployed BS density $\eta$ for different number of UEs $K$, with $\rho=20\mathrm{dB}$.}\label{Fig_simu_cell_free_near_67}
\end{figure}

\begin{figure}[!t]
    \centering
    \subfloat[Uniform distribution]
    {\begin{minipage}{1\linewidth}
        \centering
        \includegraphics[width=2.7in]{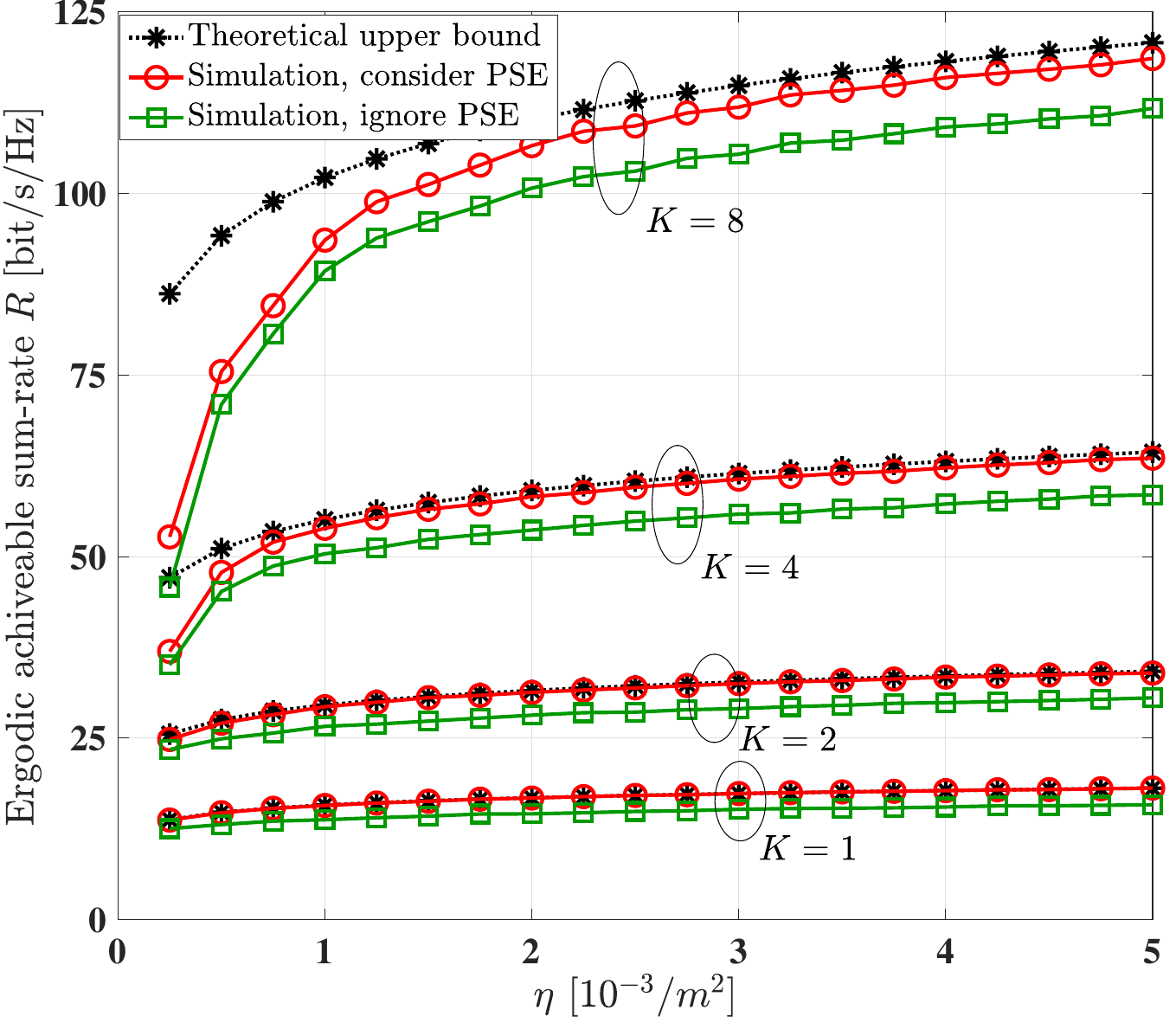}
    \end{minipage}}\\
    \subfloat[Von-Mises distribution]
    {\begin{minipage}{1\linewidth}
        \centering
        \includegraphics[width=2.7in]{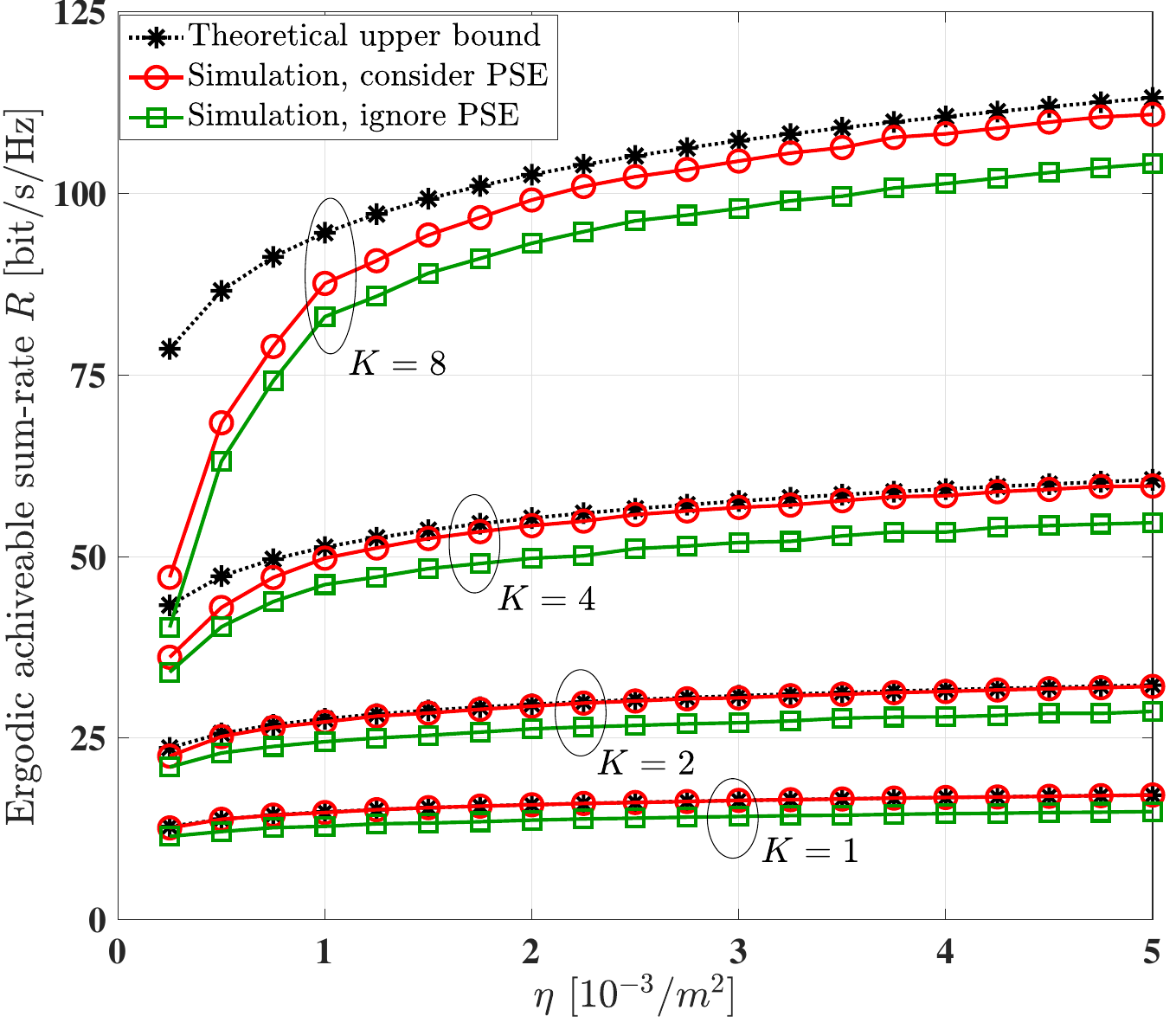}
    \end{minipage}}
    \caption{Theoretical analysis (\ref{Theoretical_Analysis_11_3}) and simulation results of the achievable sum-rate $R$ versus the deployed BS density $\eta$ for different number of UEs $K$, with $\rho=20\mathrm{dB}$ and $\sigma_\mathrm{p}^2=1$.}\label{Fig_simu_cell_free_near_89}
\end{figure}

\begin{figure}[!t]
    \centering
    \includegraphics[width=2.7in]{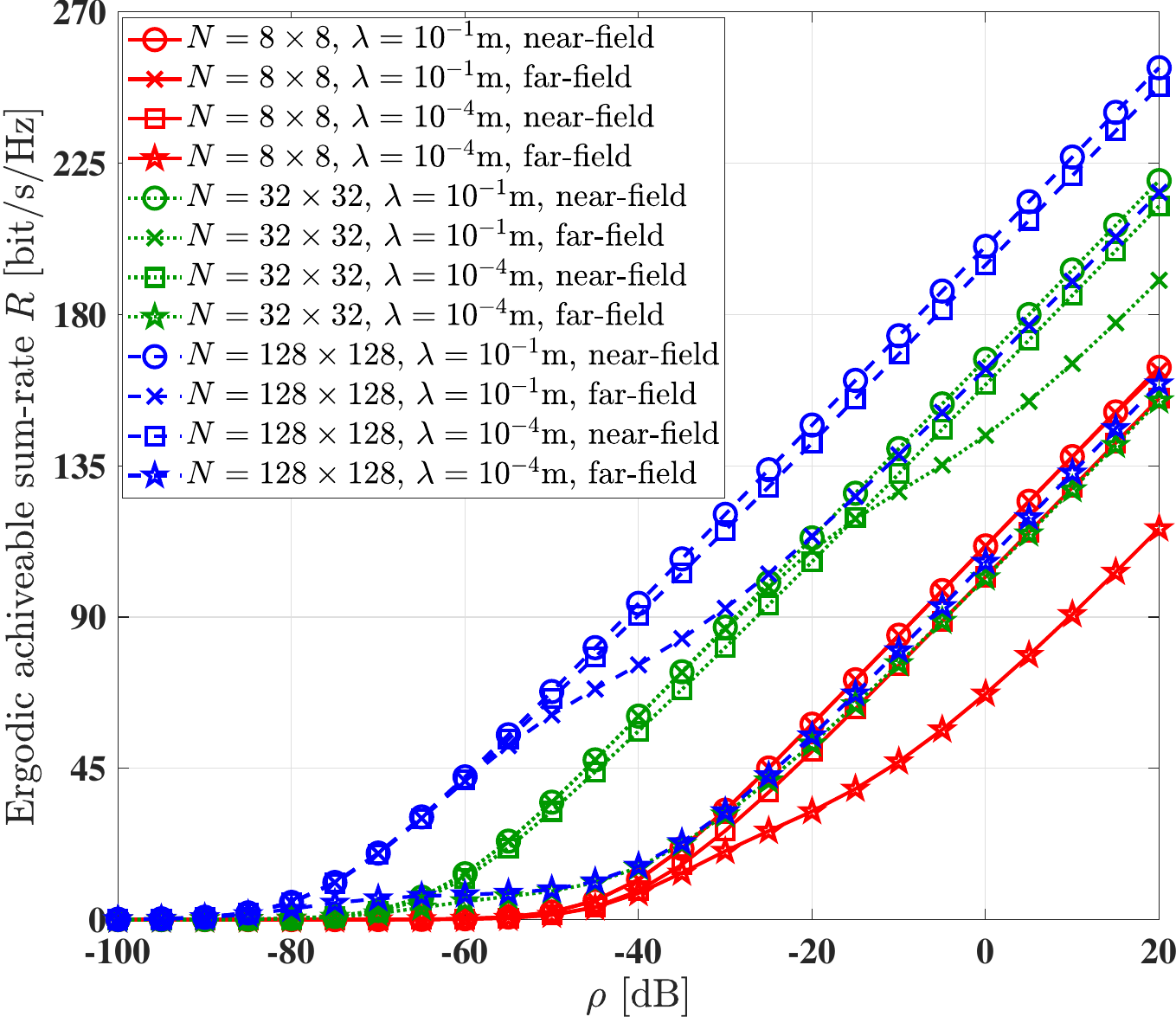}
    \caption{Simulation results of the ergodic achievable sum-rate $R$ versus the transmit power $\rho$ for different wavelength $\lambda$ and different number of RHS elements $N$, with the number of UEs $K=8$.}\label{Fig_simu_cell_free_near_10}
\end{figure}

Then, in the following we focus our attention on the multi-UE case, where $K$ UEs are randomly distributed in the coverage area. Fig.~\ref{Fig_simu_cell_free_near_67} compares the achievable sum-rate $R$ versus the BS density $\eta$ for different number of UEs, at the transmit power of $\rho=20\mathrm{dB}$. Fig.~\ref{Fig_simu_cell_free_near_67} shows that the ergodic achievable sum-rate approaches its theoretical upper bound with the increase of the BS density. Specifically, in Fig.~\ref{Fig_simu_cell_free_near_67} (a), the RF chains at the BSs have non-ideal hardware quality of $\varepsilon_v=0.99$, and it shows that the achievable sum-rate can be improved upon increasing the deployed BS density. Furthermore, the achievable sum-rate performance of the proposed beamforming method considering the HWI is better than that ignoring the HWI. By contrast, in Fig.~\ref{Fig_simu_cell_free_near_67} (b), the RF chains at the UEs have non-ideal hardware quality of $\varepsilon_u=0.99$, and it shows that the achievable sum-rate cannot be further improved by increasing the deployed BS density.

Fig.~\ref{Fig_simu_cell_free_near_89} compares the achievable sum-rate $R$ versus the deployed BS density $\eta$ with the transmit power of $\rho=20\mathrm{dB}$ and the PSE power $\sigma_\mathrm{p}^2=1$, where we show that the side effect bought by the PSE of the RHS elements can be compensated by increasing the deployed BS density. Furthermore, considering the PSE in the beamforming design can improve the achievable sum-rate.

Fig.~\ref{Fig_simu_cell_free_near_10} shows the simulation results of the ergodic achievable sum-rate $R$ versus the transmit power $\rho$ for different wavelengths $\lambda$ and different number of RHS elements $N$ with the number of UEs being $K=8$, in both the near-field and far-field channel models. The figure shows that the far-field channel model can reach similar performance as the near-field channel model, when the number of RHS elements is small i.e. $N=8\times8$, or the wavelength is large, i,e $\lambda=10^{-1}\mathrm{m}$. By contrast, the near-field channel model promises higher achievable sum-rate upon increasing the number of RHS elements $N$ or decreasing the wavelength $\lambda$. It can be illustrated that the phase shift at the receiver nodes in the far-field channel model is usually approximated by its first-order Taylor expansion based on the planar wavefront model, and this approximation results in a phase discrepancy, which increases when the distance decreases or the wavelength decreases~\cite{cui2022near}.

\section{Conclusions}\label{Conclusion}
Beamforming algorithms were formulated for reconfigurable holographic surfaces used in cell-free networks for maximizing the achievable sum-rate, in the face of realistic hardware impairments of the RF chains at the BSs and UEs as well as the phase shift error at the RHS elements. The popular stochastic geometry approach was employed for deriving the ergodic achievable sum-rate in the near-field channel model. The theoretical analysis and the simulation results showed that the achievable sum-rate is limited by the HWIs of the RF chains as well as the PSE of the RHS elements, and saturates in the high-SNR region. Furthermore, the HWIs of the RF chains at the BSs and the PSE at the RHS elements can be compensated by increasing the BSs density. By contrast, the achievable sum-rate degradation resulting from the HWIs of the RF chains at the UEs cannot be further compensated by increasing the deployed BSs density. We also showed that the ergodic achievable sum-rate considering the near-field channel model is higher than that based on the far-field channel model assumption. In our future research, the holographic beamformer of the BSs and the digital beamformer of the CPU can be jointly optimized to obtain a further performance enhancement by utilizing alternating optimization. Furthermore, both robust beamforming designs and the performance analysis of RHS based cell-free networks having imperfect CSI will be of practical interest.

\appendices
\section{Proof of Lemma~\ref{Lemma_1}}\label{Appendix_A}
Firstly, we focus on our attention the RHS phase error $\widetilde{\theta}_n^{(l)}$ following the uniform distribution, i.e. $\widetilde{\theta}_n^{(l)}\sim\mathcal{U}(-\iota_\mathrm{p},\iota_\mathrm{p})$. When $\widetilde{\theta}_n^{(l)}\sim\mathcal{U}(-\iota_\mathrm{p},\iota_\mathrm{p})$, the $i$th-order moment of $\widetilde{\theta}_n^{(l)}$, denoted as $\mathbb{E}[\widetilde{\theta}_n^{(l)i}]$, equal to 0 when $i$ is odd and equal to $\frac{1}{i+1}\iota_\mathrm{p}^i$ when $i$ is even. Thus, we arrive at
\begin{align}\label{Appendix_A_1}
    \mathbb{E}\left[\mathrm{e}^{j\widetilde{\theta}_n^{(l)}}\right]
    =\sum_{i=0}^{\infty}\frac{(-1)^i\iota_\mathrm{p}^{2i}}{(2i+1)!}
    =\frac{\sin(\iota_\mathrm{p})}{\iota_\mathrm{p}}=\xi.
\end{align}
Hence the RHS phase error power is $\sigma_\mathrm{p}^2=\mathbb{E}[\widetilde{\theta}_n^{(l)2}]=\frac{1}{3}\iota_\mathrm{p}^2$. Then, we assume that the RHS phase error $\widetilde{\theta}_n^{(l)}$ follows the von-Mises distribution, i.e. $\widetilde{\theta}_n^{(l)}\sim\mathcal{VM}(0,\varpi_\text{p})$, which satisfies~\cite{hillen2017moments}
\begin{align}\label{Appendix_A_2}
    \mathbb{E}\left[\mathrm{e}^{j\widetilde{\theta}_n^{(l)}}\right]=\frac{I_{1}(\varpi_\text{p})}
    {I_{0}(\varpi_\text{p})}=\xi.
\end{align}
Therefore, the RHS phase error power becomes $\sigma_\mathrm{p}^2=\mathbb{E}[\widetilde{\theta}_n^{(l)2}]=\frac{1}{\varpi_\text{p}}$.

According to (\ref{Channel_Model_7}), (\ref{Channel_Model_8}), (\ref{Appendix_A_1}) and (\ref{Appendix_A_2}), $\overline{h}_l^{(k)}$ can be derived in (\ref{Appendix_A_3}).
\begin{figure*}[!t]
\begin{align}\label{Appendix_A_3}
    \overline{h}_l^{(k)}=\mathbb{E}\left[\sum_{n=1}^N\sqrt{\varsigma_n^{(l)}\beta_n^{(l,k)}}
    \mathrm{e}^{\jmath\left(\overline{\theta}_n^{(l)}+\widetilde{\theta}_n^{(l)}
    -\frac{2\pi}{\lambda}\left(\left\|\mathbf{r}-\mathbf{p}_n\right\|
    +\left\|\mathbf{q}^{(l,k)}-\mathbf{p}_n\right\|\right)\right)}\right]
    =\xi\ddot{h}_l^{(k)}
\end{align}
\hrulefill
\end{figure*}
Therefore, we then get
\begin{align}\label{Appendix_A_4}
    \overline{\mathbf{h}}^{(k)}=\xi\ddot{\mathbf{h}}^{(k)}.
\end{align}
Furthermore, according to (\ref{Channel_Model_7}), (\ref{Channel_Model_9}), (\ref{Appendix_A_1}) and (\ref{Appendix_A_2}), $\mathbb{E}[h_{l_1}^{(k)}h_{l_2}^{(k)\dag}]$ can be derived in (\ref{Appendix_A_5}).
\begin{figure*}[!t]
\begin{align}\label{Appendix_A_5}
    \notag&\mathbb{E}\left[h_{l_1}^{(k)}h_{l_2}^{(k)\dag}\right]\\
    \notag=&\mathbb{E}\left[\sum_{n=1}^N\sqrt{\varsigma_n^{(l_1)}\beta_n^{(l_1,k)}}
    \mathrm{e}^{\jmath\left(\overline{\theta}_n^{(l_1)}+\widetilde{\theta}_n^{(l_1)}
    -\frac{2\pi}{\lambda}(\left\|\mathbf{r}-\mathbf{p}_n\right\|
    +\left\|\mathbf{q}^{(l_1,k)}-\mathbf{p}_n\right\|)\right)}\cdot
    \sum_{n=1}^N\sqrt{\varsigma_n^{(l_2)}\beta_n^{(l_2,k)}}\mathrm{e}^{-\jmath
    \left(\overline{\theta}_n^{(l_2)}+\widetilde{\theta}_n^{(l_2)}
    -\frac{2\pi}{\lambda}(\left\|\mathbf{r}-\mathbf{p}_n\right\|
    +\|\mathbf{q}^{(l_2,k)}-\mathbf{p}_n\|)\right)}\right]\\
    \notag=&\sum_{n_1=1}^N\sum_{n_2=1}^N\sqrt{\varsigma_{n_1}^{(l_1)}\beta_{n_1}^{(l_1,k)}}
    \mathrm{e}^{\jmath\left(\overline{\theta}_{n_1}^{(l_1)}
    -\frac{2\pi}{\lambda}\left(\left\|\mathbf{r}-\mathbf{p}_{n_1}\right\|
    +\left\|\mathbf{q}^{(l_1,k)}-\mathbf{p}_{n_2}\right\|\right)\right)}
    \sqrt{\varsigma_{n_2}^{(l_2)}\beta_{n_2}^{(l_2,k)}}
    \mathrm{e}^{-\jmath\left(\overline{\theta}_{n_2}^{(l_2)}
    -\frac{2\pi}{\lambda}\left(\left\|\mathbf{r}-\mathbf{p}_{n_2}\right\|
    +\left\|\mathbf{q}^{(l_2,k)}-\mathbf{p}_{n_2}\right\|\right)\right)}\cdot\\
    &\mathbb{E}\left[\mathrm{e}^{\jmath\left(\widetilde{\theta}_{n_1}^{(l_1)}
    -\widetilde{\theta}_{n_2}^{(l_2)}\right)}\right]
\end{align}
\hrulefill
\end{figure*}
According to (\ref{Appendix_A_1}), (\ref{Appendix_A_2}) and (\ref{Appendix_A_5}), we then get
\begin{align}\label{Appendix_A_6}
    \mathbb{E}\left[h_{l_1}^{(k)}h_{l_2}^{(k)\dag}\right]
    =\ddot{h}_{l_1}^{(k)}\ddot{h}_{l_2}^{(k)\dag},
\end{align}
if $n_1=n_2$ and $l_1=l_2$. Otherwise we have
\begin{align}\label{Appendix_A_7}
    \mathbb{E}\left[h_{l_1}^{(k)}h_{l_2}^{(k)\dag}\right]
    =\xi^2\ddot{h}_{l_1}^{(k)}\ddot{h}_{l_2}^{(k)\dag},
\end{align}
due to the independence of $\widetilde{\theta}_{n_1}^{(l_1)}$ and $\widetilde{\theta}_{n_2}^{(l_2)}$. According to (\ref{Appendix_A_6}) and (\ref{Appendix_A_7}), we can express $\mathbf{C}_{\mathbf{h}^{(k)}\mathbf{h}^{(k)}}$ as
\begin{align}\label{Appendix_A_8}
    \mathbf{C}_{\mathbf{h}^{(k)}\mathbf{h}^{(k)}}
    =\mathbb{E}\left[\mathbf{h}^{(k)}\mathbf{h}^{(k)\mathrm{H}}\right]
    =\xi^2\ddot{\mathbf{h}}^{(k)}\ddot{\mathbf{h}}^{(k)\mathrm{H}}
    +\left(1-\xi^2\right)\mathbf{Q}^{(k)}.
\end{align}
We can arrive at (\ref{Theoretical_Analysis_2}) by substituting (\ref{Appendix_A_4}) and (\ref{Appendix_A_8}) into (\ref{Beamforming_Design_5}).

\section{Proof of Theorem~\ref{Theorem_1}}\label{Appendix_B}
The SINR of $\gamma_k$ in (\ref{Theoretical_Analysis_2}) can be further reformulated as shown in (\ref{Appendix_B_1}),
\begin{figure*}[!t]
\begin{align}\label{Appendix_B_1}
    \notag\gamma_k\overset{(a)}
    \leq&\rho_k\varepsilon_u\varepsilon_v\xi^2\ddot{\mathbf{h}}^{(k)\mathrm{H}}
    \left(\rho_{k}\left(\left(1-\varepsilon_u\right)\varepsilon_v
    \xi^2\ddot{\mathbf{h}}^{(k)}\ddot{\mathbf{h}}^{(k)\mathrm{H}}
    +\left(1-\varepsilon_v\right)\xi^2\left(\ddot{\mathbf{h}}^{(k)}
    \ddot{\mathbf{h}}^{(k)\mathrm{H}}\right)
    \odot\mathbf{I}_{L}+\left(1-\xi^2\right)\mathbf{Q}^{(k)}\right)
    +\sigma_w^2\mathbf{I}_{L}\right)^{-1}\ddot{\mathbf{h}}^{(k)}\\
    \notag\overset{(b)}=&\frac{\rho_k\varepsilon_u\varepsilon_v\xi^2\mathbf{h}^{(k)\mathrm{H}}
    \mathbf{A}^{-1}\mathbf{h}^{(k)}}
    {1+\rho_k\left(1-\varepsilon_u\right)\varepsilon_v\xi^2\mathbf{h}^{(k)\mathrm{H}}
    \mathbf{A}^{-1}\mathbf{h}^{(k)}}\\
    \overset{(c)}=&\frac{\rho_k\varepsilon_u\varepsilon_v\xi^2
    \sum_{l=1}^{L}\frac{\left|h_l^{(k)}\right|^2}{\rho_k\left(1-\varepsilon_v\right)\xi^2
    \left|h_l^{(k)}\right|^2+\rho_k\left(1-\xi^2\right)\left\|\nu_l^{(k)}\right\|^2+\sigma_w^2}}
    {1+\rho_k\left(1-\varepsilon_u\right)\varepsilon_v\xi^2
    \sum_{l=1}^{L}\frac{\left|h_l^{(k)}\right|^2}{\rho_k\left(1-\varepsilon_v\right)\xi^2
    \left|h_l^{(k)}\right|^2+\rho_k\left(1-\xi^2\right)\left\|\nu_l^{(k)}\right\|^2+\sigma_w^2}}
\end{align}
\hrulefill
\end{figure*}
where we have
\begin{align}\label{Appendix_B_2}
    \notag\mathbf{A}=&\rho_k\left(1-\varepsilon_v\right)\xi^2
    \left(\ddot{\mathbf{h}}^{(k)}\ddot{\mathbf{h}}^{(k)\mathrm{H}}\right)
    \odot\mathbf{I}_{L}+\rho_k\left(1-\xi^2\right)\mathbf{Q}^{(k)}\\
    \notag&+\sigma_w^2\mathbf{I}_{L}\\
    \notag=&\mathbf{Diag}\left\{\rho_k\left(1-\varepsilon_v\right)\xi^2\left|h_1^{(k)}\right|^2
    +\rho_k\left(1-\xi^2\right)\left\|\nu_{1}^{(k)}\right\|^2\right.\\
    \notag&\left.+\sigma_w^2,\rho_k\left(1-\varepsilon_v\right)\xi^2\left|h_2^{(k)}\right|^2
    +\rho_k\left(1-\xi^2\right)\left\|\nu_{2}^{(k)}\right\|^2\right.\\
    \notag&\left.+\sigma_w^2,\cdots,\rho_k\left(1-\varepsilon_v\right)\xi^2
    \left|h_{L}^{(k)}\right|^2+\rho_k\left(1-\xi^2\right)\left\|\nu_{L}^{(k)}\right\|^2\right.\\
    &\left.+\sigma_w^2\right\},
\end{align}
with the equality in (a) established when $K=1$, (b) is based on the Woodbury matrix identity \cite{zhang2017matrix}, and (c) is based on the fact that $\mathbf{A}$ is a diagonal matrix with its inverse $\mathbf{A}^{-1}$ given in (\ref{Appendix_B_3}).
\begin{figure*}[!t]
\begin{align}\label{Appendix_B_3}
    \notag\mathbf{A}^{-1}=&\mathbf{Diag}
    \left\{\frac{1}{\rho_k\left(1-\varepsilon_v\right)\xi^2\left|h_1^{(k)}\right|^2
    +\rho_k\left(1-\xi^2\right)\left\|\nu_{1}^{(k)}\right\|^2+\sigma_w^2},
    \frac{1}{\rho_k\left(1-\varepsilon_v\right)\xi^2\left|h_2^{(k)}\right|^2
    +\rho_k\left(1-\xi^2\right)\left\|\nu_{2}^{(k)}\right\|^2+\sigma_w^2},\cdots,\right.\\
    &\left.\frac{1}{\rho_k\left(1-\varepsilon_v\right)\xi^2\left|h_{L}^{(k)}\right|^2
    +\rho_k\left(1-\xi^2\right)\left\|\nu_{L}^{(k)}\right\|^2+\sigma_w^2}\right\}
\end{align}
\hrulefill
\end{figure*}
According to (\ref{Appendix_B_1}), the ergodic sum-rate can be derived as shown in (\ref{Appendix_B_4}),
\begin{figure*}[!t]
\begin{align}\label{Appendix_B_4}
    \notag R\leq&\sum_{k=1}^K\mathbb{E}\left[\log_2\left(1+
    \frac{\rho_k\varepsilon_u\varepsilon_v\xi^2
    \sum_{l=1}^{L}\frac{\left|h_l^{(k)}\right|^2}{\rho_k\left(1-\varepsilon_v\right)\xi^2
    \left|h_l^{(k)}\right|^2+\rho_k\left(1-\xi^2\right)\left\|\nu_l^{(k)}\right\|^2+\sigma_w^2}}
    {1+\rho_k\left(1-\varepsilon_u\right)\varepsilon_v\xi^2
    \sum_{l=1}^{L}\frac{\left|h_l^{(k)}\right|^2}{\rho_k\left(1-\varepsilon_v\right)\xi^2
    \left|h_l^{(k)}\right|^2
    +\rho_k\left(1-\xi^2\right)\left\|\nu_l^{(k)}\right\|^2+\sigma_w^2}}\right)\right]\\
    \notag\overset{(a)}\leq&\sum_{k=1}^K\log_2\left(1+
    \frac{\rho_k\varepsilon_u\varepsilon_v\xi^2
    \cdot\mathbb{E}\left[\sum_{l=1}^{L}\frac{\left|h_l^{(k)}\right|^2}
    {\rho_k\left(1-\varepsilon_v\right)\xi^2\left|h_l^{(k)}\right|^2+\rho_k\left(1-\xi^2\right)
    \left\|\nu_l^{(k)}\right\|^2+\sigma_w^2}\right]}
    {1+\rho_k\left(1-\varepsilon_u\right)\varepsilon_v\xi^2
    \cdot\mathbb{E}\left[\sum_{l=1}^{L}\frac{\left|h_l^{(k)}\right|^2}
    {\rho_k\left(1-\varepsilon_v\right)\xi^2\left|h_l^{(k)}\right|^2+\rho_k\left(1-\xi^2\right)
    \left\|\nu_l^{(k)}\right\|^2+\sigma_w^2}\right]}\right)\\
    \notag\overset{(b)}=&\sum_{k=1}^K\log_2\left(1+\frac{\rho_k\varepsilon_u\varepsilon_v\xi^2
    \eta S/K\cdot\mathbb{E}\left[\frac{\left|h_l^{(k)}\right|^2}
    {\rho_k\left(1-\varepsilon_v\right)\xi^2\left|h_l^{(k)}\right|^2+\rho_k\left(1-\xi^2\right)
    \left\|\nu_l^{(k)}\right\|^2+\sigma_w^2}\right]}
    {1+\rho_k\left(1-\varepsilon_u\right)\varepsilon_v\xi^2
    \eta S/K\cdot\mathbb{E}\left[\frac{\left|h_l^{(k)}\right|^2}
    {\rho_k\left(1-\varepsilon_v\right)\xi^2\left|h_l^{(k)}\right|^2+\rho_k\left(1-\xi^2\right)
    \left\|\nu_l^{(k)}\right\|^2+\sigma_w^2}\right]}\right)\\
    \overset{(c)}\leq&\sum_{k=1}^K\log_2
    \left(1+\frac{\rho_k\varepsilon_u\varepsilon_v\xi^2\eta S/K\cdot\mathbb{E}\left[\left|h_l^{(k)}\right|^2\right]}
    {\rho_k\left(\frac{1}{\eta S}\left(1-\varepsilon_v\right)+\frac{1}{K}
    \left(1-\varepsilon_u\right)\varepsilon_v\right)\xi^2\eta S\cdot\mathbb{E}\left[\left|h_l^{(k)}\right|^2\right]+\rho_k\left(1-\xi^2\right)
    \mathbb{E}\left[\left\|\nu_l^{(k)}\right\|^2\right]+\sigma_w^2}\right)
\end{align}
\hrulefill
\end{figure*}
where (a) and (c) are based on Jensen's inequality, and (b) is based on adopting that the distribution of BSs follows a homogeneous PPP with the density of $\eta$ and each UE is allocated an average power of $\frac{\eta}{K}$. Furthermore, $\mathbb{E}[|h_l^{(k)}|^2]$ and $\mathbb{E}[\|\nu_l^{(k)}\|^2]$ can be further expressed as
\begin{align}\label{Appendix_B_5}
    \mathbb{E}\left[\left|h_l^{(k)}\right|^2\right]\leq\frac{1}{S}
    \left|\sqrt{\boldsymbol{\varsigma}}^\mathrm{H}\sqrt{\boldsymbol{\beta}^{(k)}}\right|^2,
\end{align}
and
\begin{align}\label{Appendix_B_6}
    \mathbb{E}\left[\left\|\nu_l^{(k)}\right\|^2\right]\leq\frac{1}{S}
    \boldsymbol{\varsigma}^\mathrm{H}\boldsymbol{\beta}^{(k)},
\end{align}
respectively, with $\beta_n^{(k)}$ in the vector $\boldsymbol{\beta}^{(k)}$ given by
\begin{align}\label{Appendix_B_7}
    \notag\beta_n^{(k)}
    =&\frac{1}{\pi}\iint_{\mathbf{c}\in\mathcal{C}_k}\int_{0}^{\pi}\\
    \notag&\frac{A\|\mathbf{c}\|\sin\omega}
    {4\pi(\|\mathbf{c}\|^2+2\|\mathbf{c}\|x_n\cos\omega+x_n^2+\left(H+y_n\right)^2)^\frac{3}{2}}
    \mathrm{d}\omega\mathrm{d}\mathbf{c}\\
    \notag\overset{(a)}\leq&\frac{1}{4\pi^2}\iint_{\mathbf{c}\in\mathcal{C}_o}\int_{0}^{\pi}\\
    &\frac{A\|\mathbf{c}\|\sin\omega}
    {\left(\|\mathbf{c}\|^2+2\|\mathbf{c}\|x_n\cos\omega+x_n^2
    +\left(H+y_n\right)^2\right)^\frac{3}{2}}\mathrm{d}\omega\mathrm{d}\mathbf{c},
\end{align}
where $\mathcal{C}_k$ is the BS distribution area with UE-$k$ as the coordinate system origin, the equality in (a) is established when all UEs are located at the geometrical center of the BS distribution area. According to (\ref{Appendix_B_4}), (\ref{Appendix_B_5}), (\ref{Appendix_B_6}) and (\ref{Appendix_B_7}), we can get the ergodic achievable sum-rate upper bound in (\ref{Theoretical_Analysis_4}).

\section{Proof of Theorem~\ref{Theorem_2}}\label{Appendix_C}
When we have $N_x\rightarrow\infty$ and $N_y\rightarrow\infty$, $|\sqrt{\boldsymbol{\varsigma}}^\mathrm{H}\sqrt{\boldsymbol{\beta}^{(o)}}|^2$ in (\ref{Theoretical_Analysis_4}) can be further reformulated as
\begin{align}\label{Appendix_C_1}
    \left|\sqrt{\boldsymbol{\varsigma}}^\mathrm{H}\sqrt{\boldsymbol{\beta}^{(o)}}\right|^2
    =\left|\sum_{n=1}^{\infty}\sqrt{\varsigma_n\beta^{(o)}_n}\right|^2
    \overset{(a)}=\zeta\epsilon^2,
\end{align}
where (a) is based on the fact that the amplitudes of the links spanning from the UEs to all RHS elements are approximately identical, with $\zeta$ and $\epsilon$ being
\begin{align}\label{Appendix_C_2}
    \notag\zeta=&\frac{1}{4\pi^2}\iint_{\mathbf{c}\in\mathcal{C}_o}\int_{0}^{\pi}
    \frac{A\|\mathbf{c}\|\sin\omega}{\left(\|\mathbf{c}\|^2+H^2\right)^\frac{3}{2}}
    \mathrm{d}\omega\mathrm{d}\mathbf{c}\\
    =&\frac{1}{2\pi^2}\iint_{\mathbf{c}\in\mathcal{C}_o}
    \frac{A\|\mathbf{c}\|}{\left(\|\mathbf{c}\|^2+H^2\right)^\frac{3}{2}}
    \mathrm{d}\mathbf{c},
\end{align}
and
\begin{align}\label{Appendix_C_3}
    \notag\epsilon=&\sum_{n=1}^{\infty}\sqrt{\varsigma_n}\\
    \notag=&\sum_{n=1}^{\infty}\int_{x_n-\frac{\delta_x}{2}}^{x_n+\frac{\delta_x}{2}}
    \int_{y_n-\frac{\delta_y}{2}}^{y_n+\frac{\delta_y}{2}}
    \sqrt{\frac{\left(\alpha+1\right)d_0^{\alpha+1}}{2\pi A\left(d_0^2+x^2+y^2\right)^{\frac{\alpha+3}{2}}}}\mathrm{d}x\mathrm{d}y\\
    =&\int_{-\infty}^{\infty}\int_{-\infty}^{\infty}
    \sqrt{\frac{\left(\alpha+1\right)d_0^{\alpha+1}}{2\pi A
    \left(d_0^2+x^2+y^2\right)^{\frac{\alpha+3}{2}}}}\mathrm{d}x\mathrm{d}y.
\end{align}

Furthermore, when we have $N_x\rightarrow\infty$ and $N_y\rightarrow\infty$, $\boldsymbol{\varsigma}^\mathrm{H}\boldsymbol{\beta}^{(o)}$ in (\ref{Theoretical_Analysis_4}) can be further expressed as
\begin{align}\label{Appendix_C_4}
    \boldsymbol{\varsigma}^\mathrm{H}\boldsymbol{\beta}^{(o)}
    \overset{(a)}=\zeta\sum_{n=1}^{\infty}\varsigma_n,
\end{align}
where (a) is based on the fact that the amplitudes of the links spanning from the UEs to all RHS elements are approximately identical, and $\sum_{n=1}^{\infty}\varsigma_n$ can be derived as
\begin{align}\label{Appendix_C_5}
    \notag\sum_{n=1}^{\infty}\varsigma_n=&\sum_{n=1}^{\infty}
    \int_{x_n-\frac{\delta_x}{2}}^{x_n+\frac{\delta_x}{2}}
    \int_{y_n-\frac{\delta_y}{2}}^{y_n+\frac{\delta_y}{2}}
    \frac{\left(\alpha+1\right)d_0^{\alpha+1}}{2\pi\left(d_0^2+x^2+y^2\right)^{\frac{\alpha+3}{2}}}
    \mathrm{d}x\mathrm{d}y\\
    \notag=&\int_{-\infty}^{\infty}\int_{-\infty}^{\infty}\frac{\left(\alpha+1\right)
    d_0^{\alpha+1}}{2\pi\left(d_0^2+x^2+y^2\right)^{\frac{\alpha+3}{2}}}\mathrm{d}x\mathrm{d}y\\
    =&1.
\end{align}
According to (\ref{Appendix_C_4}) and (\ref{Appendix_C_5}), we can get
\begin{align}\label{Appendix_C_6}
    \boldsymbol{\varsigma}^\mathrm{H}\boldsymbol{\beta}^{(o)}=\zeta.
\end{align}

Finally, upon putting (\ref{Appendix_C_1}) and (\ref{Appendix_C_6}) into (\ref{Theoretical_Analysis_4}), we get (\ref{Theoretical_Analysis_14}).

\bibliographystyle{IEEEtran}
\bibliography{IEEEabrv,TAMS}
\end{document}